\documentclass[12pt,preprint]{aastex}

\slugcomment{}
\shorttitle{CCD Photometry for NGC~3532}
\shortauthors{Clem et al.}

\begin{document}

\title{DEEP, WIDE-FIELD CCD PHOTOMETRY FOR \\ THE OPEN CLUSTER NGC~3532}

\author{James L. Clem and Arlo U. Landolt}

\affil{Department of Physics \& Astronomy, \\
 Louisiana State University, \\
 Baton Rouge, LA 70803-4001}

\email{jclem@phys.lsu.edu, landolt@phys.lsu.edu}

\and{}

\author{D.~W. Hoard\altaffilmark{1} and Stefanie Wachter\altaffilmark{1}}

\affil{Spitzer Science Center, \\
 California Institute of Technology, \\
 1200 E California Blvd, \\
 MC 220-6, \\
 Pasadena, CA 91125}

\email{hoard@ipac.caltech.edu, wachter@ipac.caltech.edu}

\altaffiltext{1}{Previous affiliation:  Cerro Tololo Inter-American 
Observatory, La Serena, Chile.}

\begin{abstract} 
\label{sec:abstract} 

We present the results of a deep, wide-field CCD survey for the open 
cluster NGC~3532.  Our new $BV(RI)_{c}$ photometry effectively covers a 
one square degree area and reaches an unprecedented depth of $V\sim21$ 
to reveal that NGC~3532 is a rich open cluster that harbors a large 
number of faint, low-mass stars.  We employ a number of methods to 
reduce the impact of field star contamination in the cluster 
color-magnitude diagrams, including supplementing our photometry with 
$JHK_{s}$ data from the 2MASS catalog.  These efforts allow us to 
define a robust sample of candidate main sequence stars suitable for a 
purely empirical determination of the cluster's parameters by comparing 
them to the well-established Hyades main sequence.  Our results confirm 
previous findings that NGC~3532 lies fairly near to the Sun 
[$(m-M)_0=8.46\pm0.05$; $492^{+12}_{-11}$~pc] and has an extremely low 
reddening for its location near the Galactic plane 
[$E(B-V)=0.028\pm0.006$].  Moreover, an age of $\sim300$\,Myr has been 
derived for the cluster by fitting a set of overshooting isochrones 
to the well-populated upper main-sequence.  This new photometry also 
extends faint enough to reach the cluster white dwarf sequence, as 
confirmed by our photometric recovery of eight spectroscopically 
identified members of the cluster.  Using the location of these eight 
members, along with the latest theoretical cooling tracks, we have 
identified $\sim30$ additional white dwarf stars in the $[V,~(B-V)]$ 
color-magnitude diagram that have a high probability of belonging to 
NGC~3532.  Reassuringly, the age we derive from fitting white dwarf 
isochrones to the locus of these stars, $300\pm100$\,Myr, is consistent 
with the age derived from the turnoff.  Our analysis of the photometry 
also includes an estimation of the binary star fraction, as well as a 
determination of the cluster's luminosity and mass functions.

\end{abstract}


\section{Introduction}
\label{sec:introduction}

NGC~3532 (equinox J2000.0, $\alpha=\,$11:05:33, $\delta=-$58:43:48; 
$l=289.55$, $b=+1.36$) is a beautifully expansive, very rich open 
cluster located in the constellation Carina, only three degrees from 
the star $\eta$ Carinae.  The earliest known studies of the cluster's 
characteristics include the works by \citet{Trumpler1930}, 
\citet{Wallenquist1931}, \citet{Martin1933}, and \citet{Rieke1937}.  
They determined, in order of author, the cluster's distance, 
photographic magnitudes, spectral types and distance, photovisual 
magnitudes and positions, and spectroscopic parallaxes.

More recently, there have been a few published photometric and kinematic 
investigations, based on photoelectric and/or photographic observations, 
of the cluster's properties.  The earliest known in-depth study of 
NGC~3532 was by \citet{Koelbloed1959}, who presented the first 
photoelectric photometry in the fledgling Johnson $UBV$ system for 83 
stars and derived proper motion estimates for some of the brighter stars 
in the field.  Despite the small stellar sample, Koelbloed was able to 
robustly estimate a cluster distance modulus of $(m-M)_0=8.2$ (432 pc) 
using the main-sequence fitting technique and assign values of 
$E(B-V)=0.01$ and 100\,Myr for its reddening and age, respectively, based on 
the appearance of the stars in its color magnitude diagram (CMD).

Additional photoelectric photometry for NGC~3532 in the Johnson $UBV$ 
system has been presented in the years since Koelbloed's study.  These 
include the investigations by \citet{Butler1977}, 
\citet{FernandezSalgado1980}, \citet{Johansson1981}, and 
\citet{ClariaLapasset1988}.  \citet{WizinowichGarrison1982} expanded on 
the broadband data available for this cluster by including observations 
in the Kron-Cousins $RI$ filters in addition to Johnson $UBV$.  
Moreover, a handful of photoelectric analyses of NGC~3532 has been 
presented in a variety of other niche filter systems.  Most notable of 
these are the $uvbyH\beta$ observations of \citet{Eggen1981}, 
\citet{Johansson1981}, and \citet{Schneider1987}, the data on the $DDO$ 
and Washington photometric systems presented by 
\citet{ClariaLapasset1988} and \citet{ClariaMinniti1988}, respectively, 
and Geneva photometry given by \citet{Rufener1988}.

Arguably, the most comprehensive of all the photometric investigations 
listed above (in terms of depth and sky coverage) is that of 
\citet{FernandezSalgado1980} who presented a broadband CMD for NGC~3532 
consisting of 700 stars extending down to $V\sim13.5$.  Most of their 
photometry ($\sim75\%$), however, was extracted from photographic 
plates, while the rest were obtained individually using a 
photomultiplier.  Nonetheless, their stellar sample was sufficient to 
allow them to estimate a total cluster mass of $\sim2000M_{\odot}$, 
which strongly suggests a much larger cluster population extending to 
fainter magnitudes than observed in their data.  The cluster parameters 
they derive are in good agreement with those given by 
\citet{Koelbloed1959}; namely, $(m-M)_0=8.45\pm0.27$, 
$E(B-V)=0.042\pm0.016$, and an age of $\sim200$\,Myr.

Motivated by the surprising lack of more recent photometric 
investigations for such a seemingly rich and expansive open cluster, we 
have undertaken an extensive project to obtain deep, wide-field, 
multi-epoch observations of NGC~3532.  There are two primary goals for 
these new data.  First, we wish to construct modern CCD-based CMDs for 
the cluster that extend as faint as $V\sim20$ and utilize multiple 
color indices for all stars within an approximate one square degree 
area surrounding the cluster center.  This new photometry provides a 
nearly complete census of the cluster population, including its white 
dwarfs, and allows us to more accurately derive its fundamental 
properties (i.e., distance, reddening, and age).  Secondly, using 
$V$-band images obtained at various epochs we endeavor to identify and 
characterize both the short- and long-period variable stars in the 
field, specifically those that belong to NGC~3532 itself.

While the present manuscript focuses on the new broadband $BV(RI)_{c}$ 
photometry that we have obtained for the NGC~3532 field and describes 
our efforts to better characterize its properties, a companion paper 
(in preparation) will present the results of our search for cluster 
variable stars.  An outline of the current paper is as follows.  In the 
next section we describe the observational strategy for this project 
together with a discussion of the photometric and astrometric 
calibration procedures.  Section \ref{sec:results} presents the cluster 
CMDs and compares our new, CCD-based photometry to the photoelectric 
data presented by previous studies.  We analyze these CMDs in Section 
\ref{sec:discussion} in an effort to provide new estimates for the 
cluster properties, investigate the cluster's white dwarf and binary 
population, and derive the luminosity and mass functions.  Finally, we 
conclude by summarizing our findings in Section \ref{sec:conclusions}.

\section{Observations} 
\label{sec:observations}

The current observations of NGC~3532 were collected at the Cerro Tololo 
Inter-American Observatory (CTIO) 0.9m telescope over a number of nights 
scheduled during the 2000A observing semester.  The images were obtained 
using a 2048x2048 Tektronix CCD equipped with a set of broadband 
Johnson-Kron-Cousins $BV(RI)_{c}$ filters.  This detector had a read noise of 
$5.0~e^{-}$, pixel scale of $0.401\arcsec$ pixel$^{-1}$, and field of view 
of $\sim$13.5x13.5 square arcminutes.

\begin{figure}
\plotone{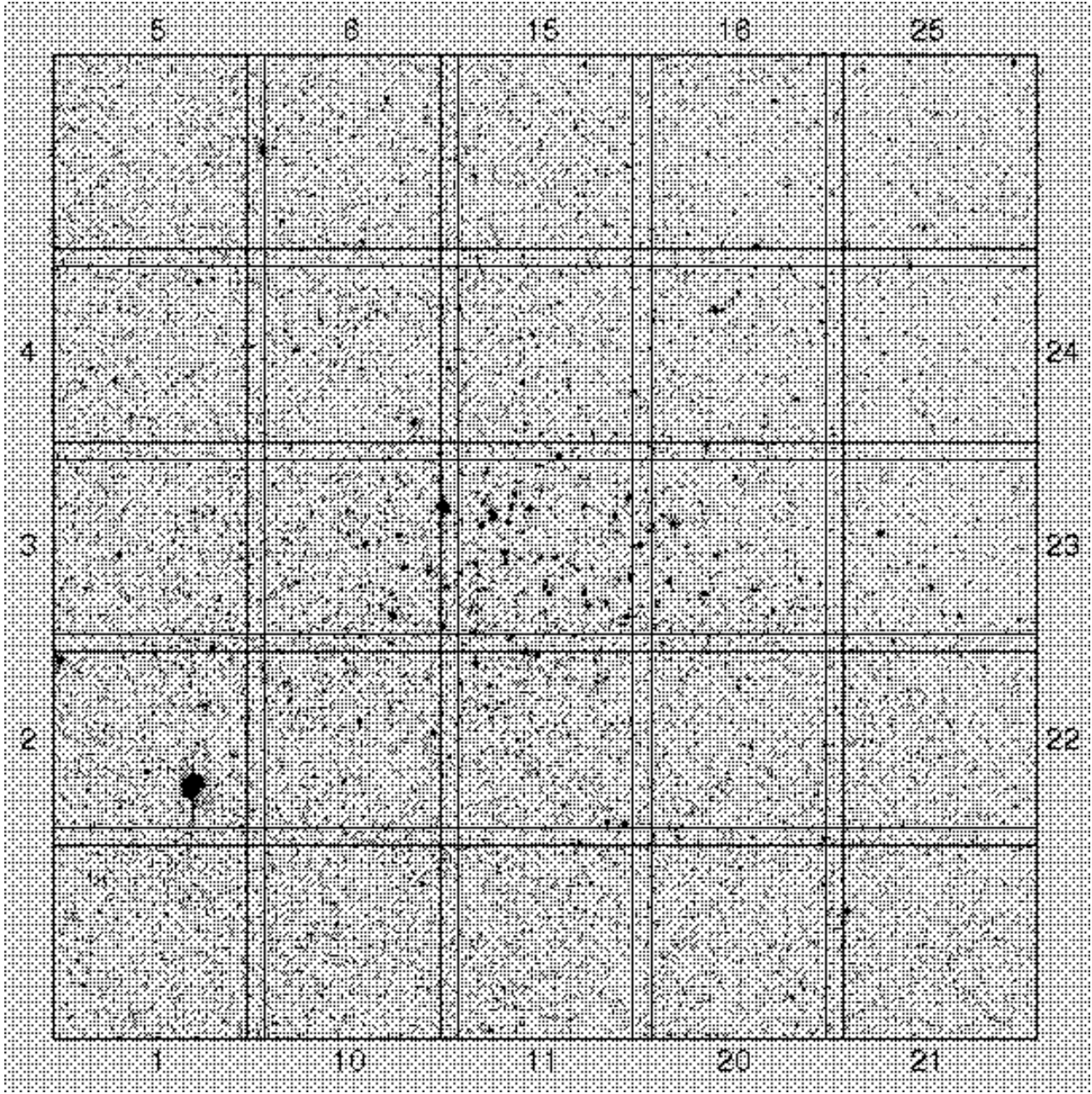}
\caption{Finding chart for the field surrounding NGC~3523 constructed from 
our best seeing $V$-band images.  North is up, and east is to the left in 
the image.  The individual squares correspond to the 13.5x13.5 arcminute 
field-of-view of the Tektronix CCD and denote the approximate locations of 
various pointings that were combined to yield a complete survey area of 
approximately 1x1 degree.}
\label{fig:fieldplot}
\end{figure}

The observing strategy for this project was designed to address the 
overall science goals mentioned in the previous section.  To obtain the 
deep, wide-field photometry of the cluster and its surrounding, our plan 
was to observe multiple fields using a varying range of exposure times in 
each of the $BV(RI)_{c}$ filters.  Note that while our program would have 
obviously benefited from the inclusion of $U$-band photometry, the 
relatively poor short-wavelength sensitivity of the CCD meant that 
performing observations in the $U$ filter would have required a 
significant fraction of our allocated telescope time.  Given the 
effective field of view for the Tektronix CCD, we defined a grid pattern 
of 25 different telescope pointings, centered on $\alpha=\,$11:05:37, 
$\delta=-$58:43:07 (J2000.0), that would combine to achieve a total one 
square degree sky coverage.  A finding chart for the NGC~3532 field, 
constructed from our best seeing CCD images, that illustrates the pattern 
of our observations, is given in Figure \ref{fig:fieldplot}.  Note that 
each of these fields overlapped its neighbors by at least one arcminute 
to allow comparison of photometry derived for adjacent pointings.  Since 
our desire was to obtain high-precision $BV(RI)_{c}$ photometry for the 
majority of stars in the cluster extending much fainter than previous 
studies, we tailored the exposure times to achieve good signal-to-noise 
ratios for cluster stars ranging in magnitude between $V=5-20$.  
Therefore, each of our 25 fields was observed using exposure times of 1s, 
5s, 30s, 100s, and 240s.  Ultimately, we were able to obtain at least one 
observation in $B$, $V$, $R_{c}$, and $I_{c}$ using each of these 
exposure times for every field.  Moreover, during photometric nights, a 
number of standard star fields from the lists of \citet{Landolt1992} were 
also observed to facilitate the transformation of our instrumental 
photometry to the standard system.

To address the secondary goal of identifying and characterizing the 
variable stars in the field, our observing runs were scheduled such 
that $V$-band observations could be collected over the entire 6 month 
period from 2000 Feb-Jul to help identify both short- and long-term 
variables.  An outline of our observing runs is listed in Table 
\ref{tab:table1}, where the number of frames collected in each filter, 
along with the fields observed in NGC~3532 (c.f., Figure 
\ref{fig:fieldplot}), is provided.  Note that the run scheduled for 
2000 Feb 25 - Mar 1 was devoted primarily to obtaining the cluster 
$BV(RI)_{c}$ photometry and standard stars were observed depending on 
photometric conditions.  For the nights of 2000 Mar 17-21, we observed 
all 25 of the cluster fields in the $V$-band with the aim of obtaining 
at least 1 epoch of observations per field per night.  Finally, the 
scattering of nights scheduled in 2000 Feb, Mar, Apr, May, and Jun took 
advantage of the CTIO Synoptic, Sidereal, and Target of Opportunity 
(SSTO) pilot program to collect additional $V$-band observations of the 
cluster at a variety of epochs spanning the observing semester to help 
pick out long-term variables.  Due to the limitations of the SSTO 
program, however, we were limited to observing no more than 3 fields 
per night. Hence, we identified 3 new subfields, designated fields 26, 
27, and 28, that approximately overlap fields 13, 14, and 15, 
respectively, in Figure \ref{fig:fieldplot} and were selected to 
provide a radial sampling of the cluster stars.  It is important to 
note that exposure times used for our cluster monitoring were identical 
to those listed above, namely, each field was observed using 
integrations of 1s, 5s, 30s, 100s, and 240s.

Overall, the observing program netted a total of 2171 frames for the 
NGC~3532 field (150 frames each in $B(RI)_{c}$ and 1721 frames in $V$). 
Moreover, we collected 232 observations of Landolt standard fields during 
photometric occasions on 5 separate nights (60 in $B$, 64 in $V$, and 54 
in both $R_{c}$ and $I_{c}$).  Ultimately, the $BV(RI)_{c}$ observations 
that were obtained on these nights are used to construct the cluster CMDs 
that will be analyzed here, while the $V$-band frames taken over the 
entire 6 month period will be used in the companion paper to search for 
variable stars.

All cluster and standard frames were preprocessed (i.e., bias 
subtraction and flat fielding) using the standard set of tasks within 
IRAF\footnote{IRAF is distributed by the National Optical Astronomy 
Observatories, which is operated by the Association of Universities for 
Research in Astronomy, Inc., under contract to the National Science 
Foundation.}.  Once instrumental signatures were removed from these 
frames, the analysis moved to extracting photometry for both the 
cluster and standard stars.  For the latter, we relied upon standard 
aperture photometry using an aperture diameter of $14\arcsec$ since 
this was the same size favored by \citet{Landolt1992, Landolt2009} in 
measuring the flux of his standard stars.  For stars in the cluster 
fields, on the other hand, we employed the technique of PSF-fitting as 
part of the DAOPHOT/ALLSTAR suite of algorithms \citep{Stetson1987, 
StetsonHarris1988} to derive their instrumental magnitudes.  Briefly, 
this technique involved detecting star-like objects in the image, 
constructing a model PSF from a subsample of these objects, and 
subsequently subtracting this model from the detections to yield a list 
of instrumental magnitudes, associated uncertainties, and $(x, y)$ 
positional information for stars in the frame.  This technique was 
iterated upon 2-3 times using between 50 and 150 stars in each image 
(depending on exposure time) to construct the model point spread 
function.  During each iteration, the subtracted image was fed back 
into the star detection algorithm to locate stars that might have been 
missed in previous runs due to the effects of crowding.

Once the instrumental photometry had been extracted from each image, it 
was necessary to place the PSF-based magnitudes on a more absolute, 
aperture-based scale similar to those derived for the standard stars.  
This step was accomplished using the method of concentric aperture 
growth-curve analysis as part of the DAOGROW package 
\citep{Stetson1990} with the goal to derive a set of ``aperture 
corrections'' which, when applied to the PSF-determined magnitudes, 
allowed the measurements made on different nights and during different 
seeing conditions to be placed on a homogeneous, aperture-based system.  
The resulting aperture-corrected PSF magnitudes can be readily 
transformed to the standard $BV(RI)_{c}$ system defined by the standard 
stars by employing typical calibration techniques.

In order to translate the CCD-based $(x, y)$ positions derived for the 
cluster and standard stars to a more meaningful $(\alpha, \delta)$-based 
system, we employed the positional information given in the UCAC3 catalog 
\citep{Zacharias2010} as our astrometric reference.  In brief, the UCAC3 
catalog positions are used to derive a set of preliminary transformation 
equations that account for offset, scale, and rotation differences between 
the CCD-based and astrometric standard systems by relying on only a 
handful of the brightest stars in each field.  Next, these initial 
transformation estimates are fed into the DAOMASTER task in an effort to 
improve their precision and accuracy by employing a set of third-order 
polynomials that further account for small higher-order effects (e.g., due 
to optical distortions, filter-induced scale differences, and/or 
differential refraction) in the CCD images.  These transformations, which 
are based on a larger sample of stars in common between the CCD image and 
the UCAC3 astrometric reference catalog, are iterated upon until a 
matching tolerance of $\sim0.5\arcsec$ is achieved and the total number of 
stars in the master list stabilizes.  The average RMS of the residuals 
resulting from the fitting process generally ranged between 
$0.1-0.3\arcsec$ for each frame, which is in agreement with the 
characteristic accuracy of the positions in the UCAC3 catalog.  In the 
end, this technique allowed us to derive a master $(\alpha, \delta)$-based 
list of detected objects for not only the cluster field, but also the 
fields containing the standard stars to help facilitate the matching of 
the photometry from frame-to-frame.

Once the master star list for NGC~3532 had been created and the 
astrometric transformation equations determined using these methods, 
the entire set of 2171 cluster frames, along with their associated PSF 
models and photometry lists, are were into the ALLFRAME program 
\citep{Stetson1994} in an effort to improve upon the profile-fitted 
photometry by deriving a homogeneous set of positions and magnitudes 
for all detected objects regardless of seeing conditions, crowding 
effects, or filters employed.

To transform our instrumental photometry to the standard system, 
extinction, zero-point and color transformation coefficients were derived 
using iterative weighted least squares fits to the photometry for 60-80 
standard stars, depending on filter.  These fits also reject stars with 
large residuals to ensure spurious measurements do not influence the 
overall quality of the transformations.  Equations of the form

$B_{obs}=B_{std}+X_{B}+C_{B}(B-V)_{std}+Z_{B}$,

$V_{obs}=V_{std}+X_{V}+C_{V}(B-V)_{std}+Z_{V}$,

$R_{obs}=R_{std}+X_{R}+C_{R}(V-R)_{std}+Z_{R}$, and

$I_{obs}=I_{std}+X_{I}+C_{I}(V-I)_{std}+Z_{I}$

\noindent{were used to transform the photometry, where the terms on the 
left sides of the equations represent the observed instrumental 
magnitudes, and those on the right are their corresponding values on 
the standard system.  The $X$, $C$, and $Z$ terms represent the 
extinction, color, and zero-point coefficients, respectively, that were 
each calculated on a nightly basis when standard stars were observed.}

\begin{figure}
\plotone{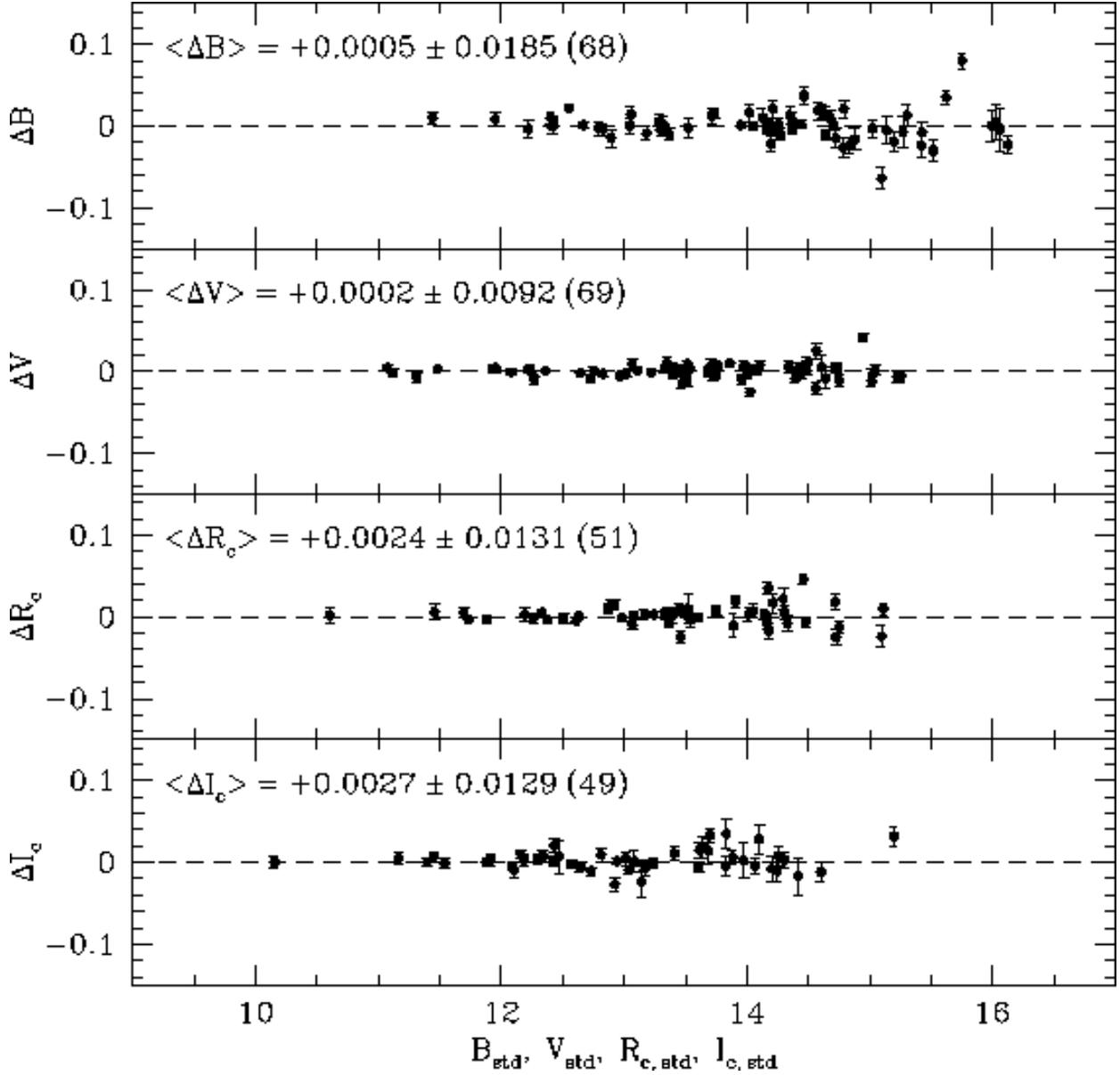}
\caption{Comparison of the $BV(RI)_{c}$ magnitudes recovered from our 
photometric calibrations with those published by \citet{Landolt2009} 
for the standard stars that were observed in our program.  Each 
$\Delta$mag is plotted as a function of its corresponding magnitude and 
is in the sense of our photometry minus Landolt's.  Dashed horizontal 
lines mark the location of zero difference.  The computed mean 
differences, standard deviations, and number of stars are given in each 
panel.}
\label{fig:compstdmag}
\end{figure}

\begin{figure}
\plotone{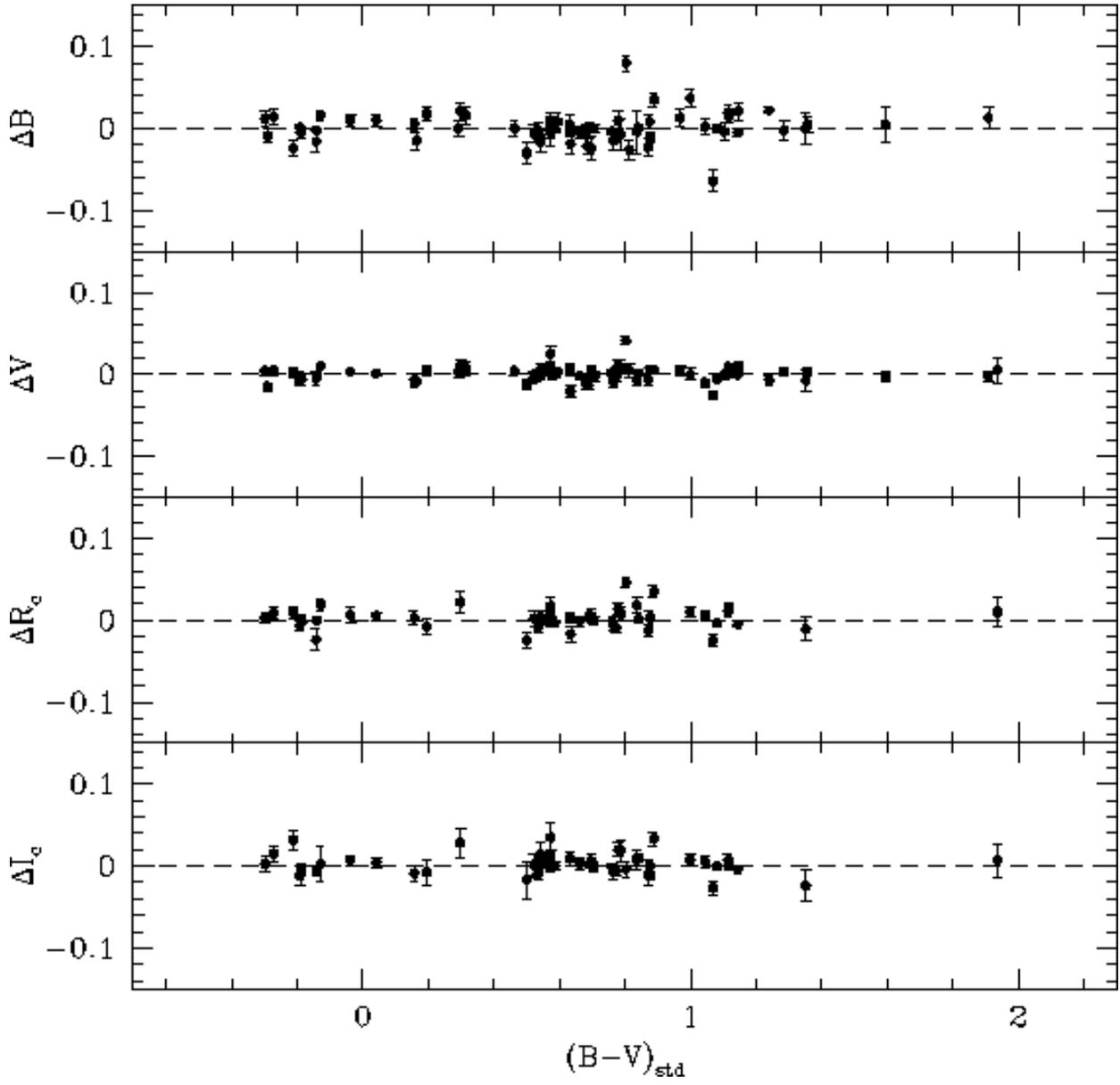}
\caption{Same as Figure \ref{fig:compstdmag} except plotted as a function 
of the standard $(B-V)$ colors given by \citet{Landolt2009}.}
\label{fig:compstdcol}
\end{figure}

The overall quality of these photometric transforms can be tested by 
comparing the computed magnitudes of our observed standard stars to 
their counterparts on the standard system.  Figures 
\ref{fig:compstdmag} and \ref{fig:compstdcol} show such comparisons, 
with the differences in the individual magnitudes, in the sense of ours 
minus Landolt, plotted versus standard magnitude and standard $(B-V)$ 
color, respectively.  In each plot, we limit the number of stars based 
on the conditions that each must have a photometric uncertainty less 
than $0.03$ mag (standard error of the mean) and be measured at least 2 
times in both data sets.  Plots such as these help to identify possible 
trends in the residuals that would warrant the inclusion of additional 
terms in the transformation equations.  Reassuringly, there appear to 
be no strong systematic differences between our recovered magnitudes 
and those of \citet{Landolt2009}, and the majority of data points 
cluster quite tightly about the loci of zero photometric difference, 
denoted as dashed horizontal lines.  Based on the mean magnitude 
differences indicated in each panel of Figure \ref{fig:compstdmag}, we 
deduce that the zero-points of our NGC~3532 photometry are accurate to 
$\sim0.003$ mag or better for any given filter.

Once the photometric coefficients were determined, it was simply a matter 
of applying the transformation equations in reverse to yield calibrated 
photometry for the cluster stars.  For this step, the ALLFRAME-determined, 
aperture-corrected PSF magnitudes for a small sample of bright stars in 
each image, typically those selected to construct the PSF, were first 
transformed to the standard system to serve as a set of local secondary 
standards.  The calibrated photometry for this subsample is subsequently 
used to relate the remaining stars in each field to the standard system by 
determining frame-to-frame zero-point differences that may exist due to 
short-term variations in extinction and/or errors in the aperture 
corrections.

\section{Results}
\label{sec:results}

\subsection{Current Photometry}
\label{subsec:currentphot}

Our reductions of the entire NGC~3532 field netted both astrometric 
positions and $BV(RI)_{c}$ photometry for 316,367 objects within a one 
square degree area surrounding the cluster center.  An example of these 
data is given in Table \ref{tab:table2}.  From this sample, we will 
consider a total of 285,990 objects for further analysis since they 
were detected at least once in each of the $B$, $V$, $R_{c}$, and 
$I_{c}$ filters.  The fact that the vast majority of the remaining 
30,377 excluded objects ($\sim93\%$) have $V\gtrsim20$ implies they 
likely went undetected in one or more filters due to their extreme 
faintness and/or the effects of crowding.

\begin{figure}
\plotone{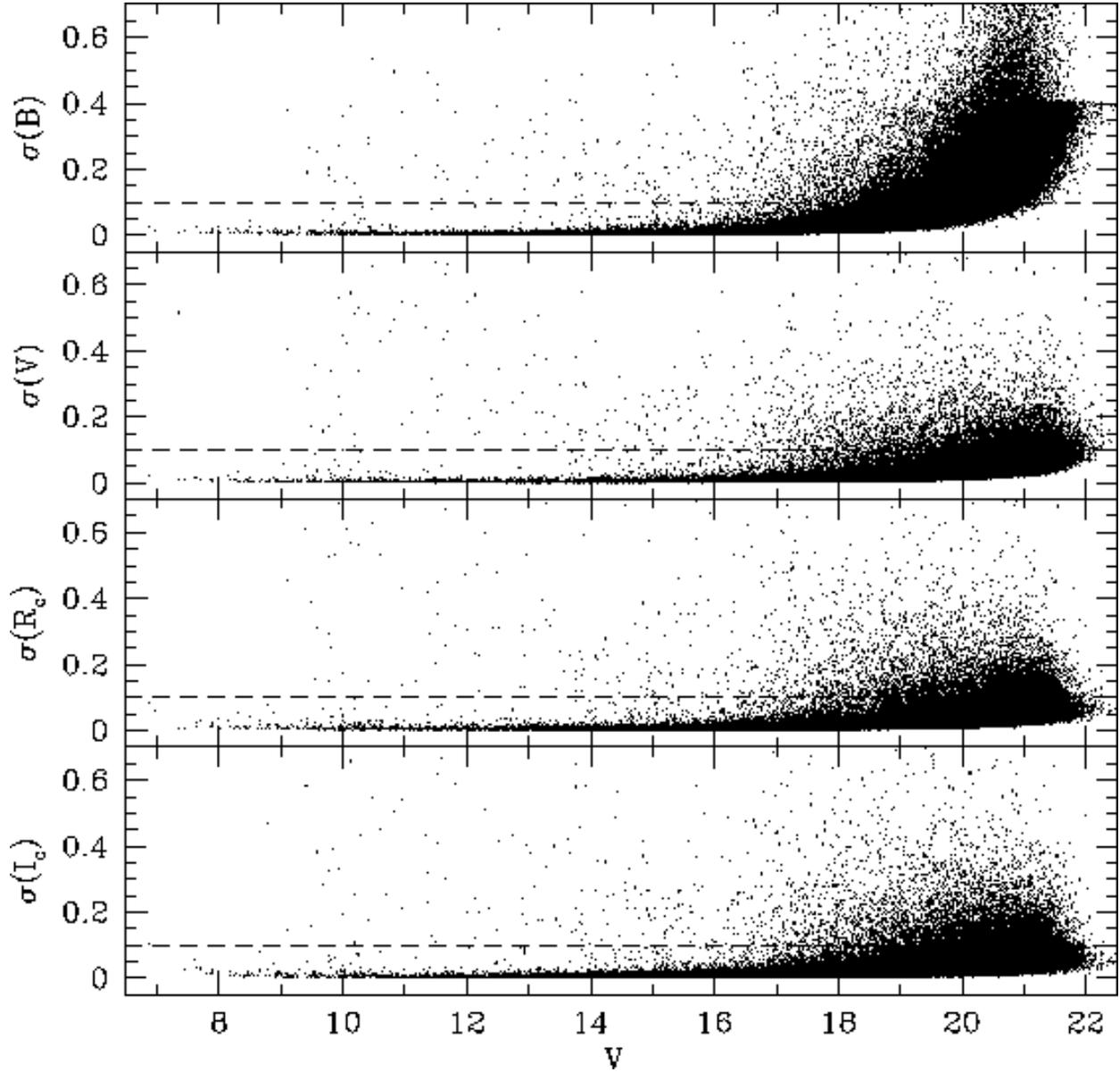}
\caption{The uncertainties in our derived $BV(RI)_{c}$ photometry 
(standard error of the mean magnitude) plotted as a function of $V$ 
magnitude.  Stars lying below the dashed horizontal lines, corresponding 
to $\sigma(mag)=0.1$, are those deemed to have the highest quality 
photometry for our analysis.}
\label{fig:photerr}
\end{figure}

In Figure \ref{fig:photerr}, we plot the uncertainties in each 
magnitude, $\sigma(mag)$, versus $V$ for the sample of objects with 
valid measurements in all four filters.  These uncertainty values 
represent the standard error of the mean magnitude as computed by our 
reductions.  Depending on the number of measurements for a particular 
object, $\sigma(mag)$ is dominated by either the internal noise 
estimates in the photometric reductions or by the external 
frame-to-frame agreement in the calibrated magnitudes.  Note the high 
precision in our photometry, with the majority of stars having 
$\sigma(mag)\lesssim0.1$ for the $V$, $R_{c}$, and $I_{c}$ filters 
over the entire magnitude range.  The $B$-band uncertainties, on the 
other hand, begin to rise above 0.1 mag at $V\sim19$, indicating that 
our observing program would have benefited from longer exposure times 
in $B$ in order to achieve the same depth as the other filters.

\begin{figure}
\plotone{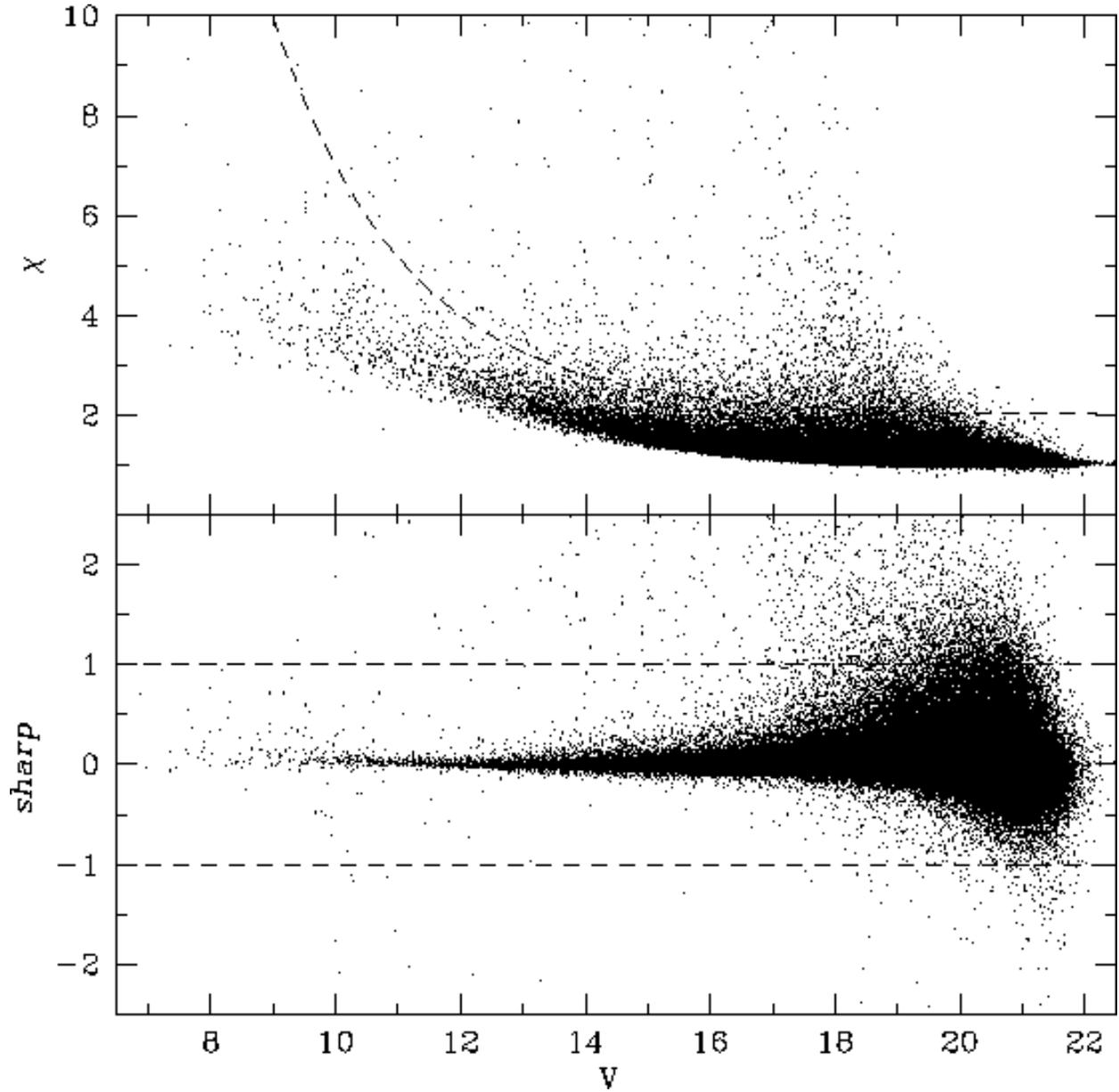}
\caption{The image quality statistics, $\chi$ and $sharp$, plotted as a 
function of $V$ magnitude.  The dashed lines denote the cuts we have used 
to exclude objects that might have spurious photometry due to image 
defects or are background galaxies.}
\label{fig:photparams}
\end{figure}

The ALLFRAME reductions also supply two image quality parameters known 
as $\chi$ and $sharp$ that are based on the pixel-to-pixel residuals 
between the model PSF and the observed brightness profile for any given 
object \citep{Stetson2003}.  While the former can be used to separate 
out objects that are contaminated by image defects, bad pixels or 
diffraction spikes, the latter is useful for isolating legitimate stars 
from background galaxies.  The $\chi$ and $sharp$ estimates given in 
Table \ref{tab:table2} for any given entry represent the mean of those 
determined individually for each frame in which that object was 
detected.  Figure \ref{fig:photparams} provides a plot of the $\chi$ 
and $sharp$ values versus $V$ magnitude for the same number of objects 
shown in Figure \ref{fig:photerr}.

The $[V,~(B-V)]$ and $[V,~(V-I_{c})]$ color-magnitude diagrams (CMDs) for 
NGC~3532 that result from our reductions are shown in Figure 
\ref{fig:cmds_init}.  In both panels we plot only those stars judged to 
have the highest quality photometry based on their photometric 
uncertainties, the number of independent measurements in each filter, and 
values of the ALLFRAME-computed image quality statistics, $\chi$ and 
$sharp$.  Specifically, we have plotted stars that have at least one 
measurement in any given filter together with $\sigma(mag)\leq0.1$ mag, 
$\chi\leq2.0+10^{-0.2(V-13.5)}$, and $|sharp|\leq1.0$ (we have denoted 
these limits by the dashed lines in Figures \ref{fig:photerr} and 
\ref{fig:photparams}).

Upon inspection of the CMDs in Figure \ref{fig:cmds_init}, a few things 
are immediately evident.  First, a well-defined cluster main sequence 
can clearly be seen extending from the turnoff at $V\sim8$ down to 
$V\sim16$ where it begins to become lost in field star contamination.  
The significant number of field stars in the CMDs undoubtedly arises 
due to the low Galactic latitude of the cluster.  Within the field star 
distribution there appear to be two separate populations; one 
corresponding to the field dwarfs that lies fainter and blueward of the 
cluster main sequence, with a second distinct population associated 
with field giant stars that can be identified as the plume of stars at 
$(B-V)$ and $(V-I_{c})\sim1.5$.  These field stars begin to overlap the 
cluster population at $V>16$, making it quite difficult to ascertain 
the exact location of the cluster main sequence at fainter magnitudes.

At the brightest end of each CMD there is a handful of cluster red 
giants at $(B-V)$ and $(V-I_{c})\sim1$.  Such a small population of giant 
stars, combined with the near-vertical nature of the turnoff and upper 
main sequence points to the fact that NGC~3532 is a fairly young 
cluster (i.e., $<1$\,Gyr).  In addition, there are two objects lying 
between the turnoff and giant clump at $V\sim8$ that may be cluster 
stars transitioning the Hertzsprung gap.  A more detailed analysis of 
the giant star population in NGC~3532 will be given in Section 
\ref{subsec:giants}.

Finally, the $[V,~(B-V)]$ CMD in the left-hand panel of Figure 
\ref{fig:cmds_init} also shows the presence of several faint blue objects 
that we presume to be part of the white dwarf sequence of NGC~3532.  
Moreover, it appears that the majority of these stars are positioned in a 
clump lying at $V\sim$20 and $(B-V)\sim0$, with only a handful of stars 
extending fainter.  White dwarfs as faint as $V\sim20$ are known to exist 
in NGC~3532 based on the investigations of \citet{ReimersKoester1989} and 
\citet{KoesterReimers1993}.  It may indeed be the case that this clump 
corresponds to the end of the white dwarf cooling sequence.  If so, it 
would provide an independent method for deriving the cluster age by 
fitting white dwarf cooling models to the observed population.  A more 
detailed analysis of these faint, blue objects will be given in Section 
\ref{subsec:wds}.

\subsection{Comparisons with Previous Photometry}
\label{subsec:compphot}

A number of other photometric studies have been published for NGC~3532, 
though none that provide the depth and coverage of this one.  Using the 
excellent resources available at the WEBDA 
website\footnote{\url{http://www.univie.ac.at/webda/}}, we have been 
able to cross-identify stars that are in common between our study and 
the WEBDA database for the purpose of comparing the broadband 
photometry presented by previous studies for NGC~3532.  Specifically, 
we consider the photometric data given in the following publications: 
\citet[][hereafter K59]{Koelbloed1959} (82 stars), \citet[][hereafter 
B77]{Butler1977} (26 stars), \citet[][hereafter 
FS80]{FernandezSalgado1980} (180 stars), \citet[][hereafter 
J81]{Johansson1981} (14 stars), \citet[][hereafter 
WG82]{WizinowichGarrison1982} (68 stars), and \citet[][hereafter 
CL88]{ClariaLapasset1988} (12 stars).  All of these investigations 
provided Johnson $UBV$ photometry for their stars, while the work of 
WG82 is the only one to include Cousins $RI$ photometry.

Table \ref{tab:table3} presents the correspondence between the WEBDA 
identification system for stars given in these studies and our own 
numbering system, along with the equatorial coordinates and photometric 
information derived as part of our analysis.  In Table \ref{tab:table4} we 
have listed the previously published photometry from the studies listed 
above for all stars in Table \ref{tab:table3}.  It should be noted that 
due to space limitations, Table \ref{tab:table4} only includes the data 
relevant to our comparisons, and excludes any information that has little 
or no impact on the analysis presented here (e.g., some of these studies 
provide $(U-B)$ colors, number of individual measurements for each star, 
magnitude and/or color uncertainties, etc.).

By combining the information listed in both Tables \ref{tab:table3} and 
\ref{tab:table4}, we are able to create Figures \ref{fig:mkcompmag} and 
\ref{fig:mkcompcolor} that show the differences in $V$ magnitudes and 
$(B-V)$ colors, respectively, for each of the six previous photometric 
studies of NGC~3532.  In both figures, the magnitude and color 
differences plotted along the ordinates are in the sense of the 
indicated study minus our present data, while the abscissae give our 
$V$ magnitudes and $(B-V)$ colors.  Some of the panels include open 
triangles that are meant to point to data lying beyond the ranges of 
these plots, and a few of these triangles are labeled with their WEBDA 
identification numbers for further discussion below.

An examination of our finding chart in Figure \ref{fig:fieldplot} reveals 
that WEBDA stars \#426, \#262, and \#255 are situated very close to other 
bright stars; thus, the $V$ magnitudes presented by the FS80 study are 
likely too bright when compared with ours since their measuring aperture 
probably contained too much light from the neighbors.  This assumption is 
supported by our comparisons when considering that the $\Delta V$ values 
for \#426, \#262, and \#255 are all negative ($\Delta V=-0.705$, $-0.357$, 
$-0.545$, respectively).  However, we cannot rule out that these stars may 
actually exhibit some variability that is the cause of such large $\Delta 
V$ values.

In addition to the discussions of these individual stars, we also point 
to the large scatter in the comparison of our $V$ magnitudes to those 
of B77 for stars fainter than $V\sim16$ in panel (b) of Figure 
\ref{fig:mkcompmag}.  Indeed, B77 claims that his photometry for 
stars with $V>16$ has a high degree of uncertainty among their 
individual measurements due to their extreme faintness.  Given this, 
along with the large scatter seen at the faint end of panel (b), we do 
not place any significance on the B77 photometry for stars with $V>16$.

Furthermore, comparisons involving the $(RI)_{c}$ photometry of WG82 
exhibit some peculiarities as shown in Figure \ref{fig:mkcompWG}.  
There are strong systematic differences in both $(V-R_{c})$ and $(R-I)_{c}$ 
between our photometry and theirs as shown in the top row of the 
figure.  Unfortunately, no other studies of NGC~3532 using the Cousins 
$RI$ filters exist in the literature that would help to identify which 
set of photometry is at fault.  Based on the fact, however, that 
Figures \ref{fig:compstdmag} and \ref{fig:compstdcol} of the present 
investigation show that the recovered $(RI)_{c}$ magnitudes of our 
observed standards are in very good agreement with those given by 
\citet{Landolt2009}, we conclude that the problem actually lies in the 
$(RI)_{c}$ photometry tabulated by WG82.  Rather than delve too deeply 
into this, we simply assume that these systematic differences can be 
corrected using a simple linear relationship.  Based on a least-squares 
fit to the data we find the slopes of the lines shown in the top panels 
of Figure \ref{fig:mkcompWG} to be $0.391$ for $(V-R_{c})$ and $0.143$ for 
$(R-I)_{c}$.  Using this information to correct the WG82 photometry results 
in the plots shown in the bottom panels.

Table \ref{tab:table5} gives our computed differences in the $V$ 
magnitudes and colors for the indicated studies.  The ``Clipped Mean" 
column represents a determination of the mean based on an iterative 
clipping scheme where all stars that lie beyond 3 times the standard 
deviation are rejected from the computation of the average.  The stars 
removed in this way are indicated in both Figures \ref{fig:mkcompmag} 
and \ref{fig:mkcompcolor} as open circles or open triangles.  The 
number of stars used in each of these computations is also provided 
along with the number of stars rejected.  Note that our mean values 
derived for the photometry in the B77 investigation include stars 
brighter than $V=16$.  Also note that the means given for the 
$(V-R_{c})$ and $(R-I)_{c}$ colors of WG82 assume that the systematics 
shown in the top panels of Figure \ref{fig:mkcompWG} have been removed.

Among the studies considered here, the one exhibiting the largest 
differences, both in $V$ magnitude and $(B-V)$ color, is that of WG82.  
Indeed, WG82 were aware of a zero-point difference between their data when 
compared to both K59 and FS80, despite all three of these investigations 
using E-region standards to calibrate their photometry.  They cited 
declination effects in their observing equipment as a possible explanation 
for these differences.  Whatever the reason, our value of $\Delta 
V\sim+0.07$ for their study is the largest of the six.  This fact, 
together with the earlier discussion regarding the strange systematics in 
their $(V-R_{c})$ and $(R-I)_{c}$ colors (c.f., Figure \ref{fig:mkcompWG}), is an 
indication that their photometry for NGC~3532 should be used with caution.

\subsection{Isolating the Main Sequence}
\label{subsec:ms}

\begin{figure}
\plotone{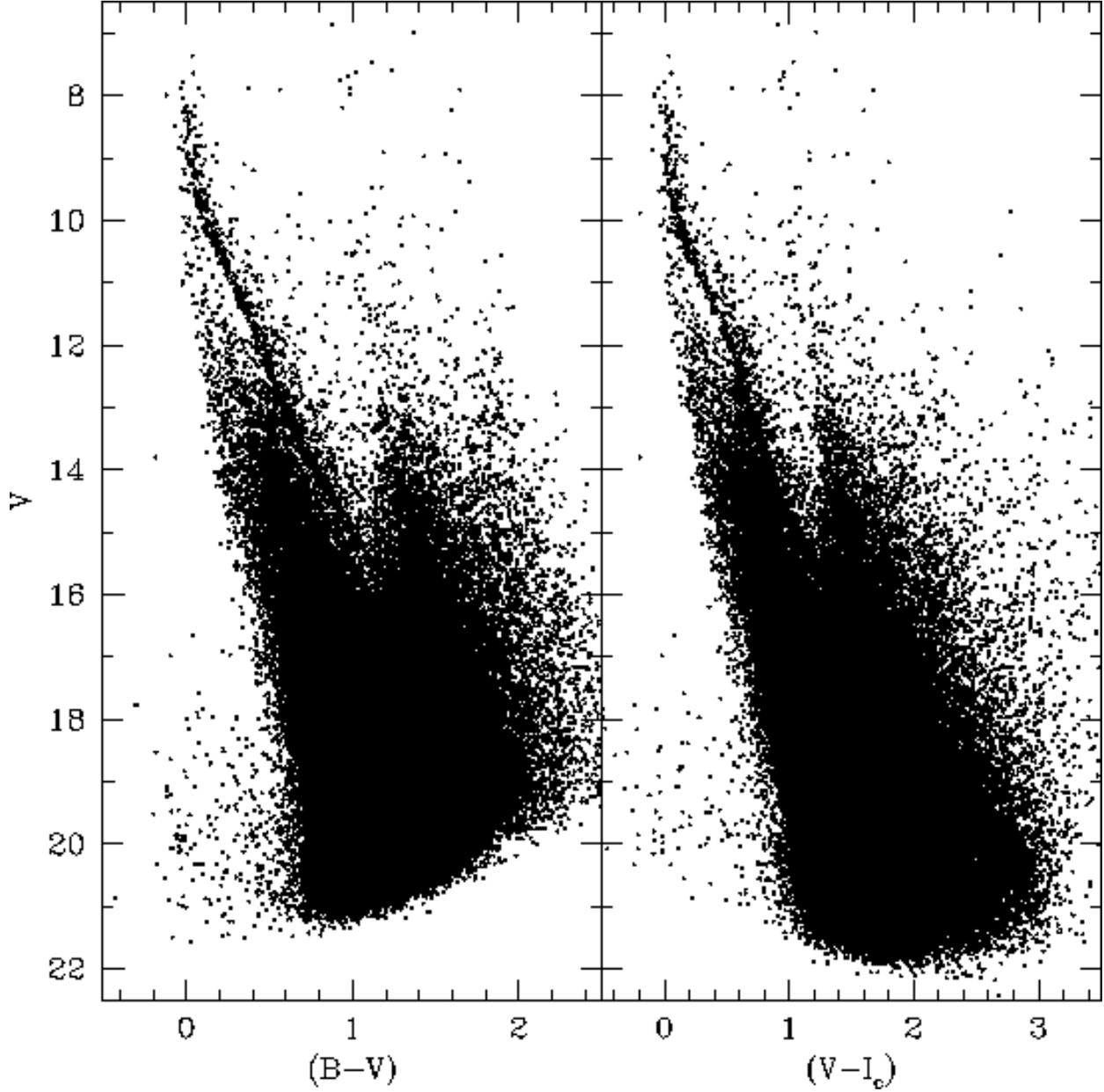}
\caption{Color-magnitude diagrams for the sample of objects in the 
NGC~3532 field judged to have the best-quality photometry based on their 
photometric uncertainties and $\chi$ and $sharp$ values as described in 
the text and illustrated in Figures \ref{fig:photerr} and 
\ref{fig:photparams}.}
\label{fig:cmds_init}
\end{figure}

\begin{figure}
\plotone{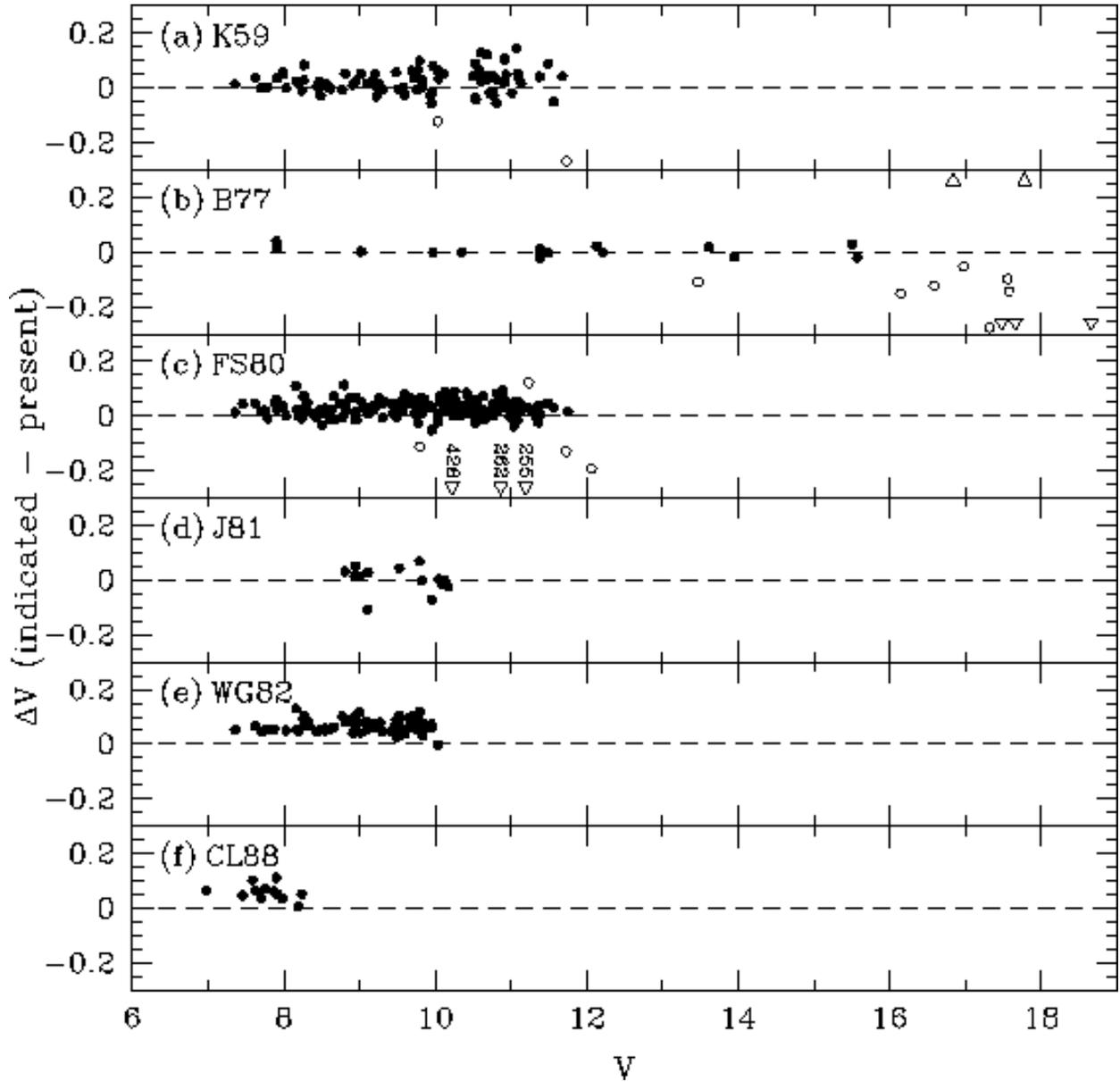}
\caption{Comparisons of the $V$-band photometry derived by the various 
indicated studies with ours.  Dashed horizontal lines correspond to zero 
difference, and the meaning of the different symbols is described in the 
text.}
\label{fig:mkcompmag}
\end{figure}

\begin{figure}
\plotone{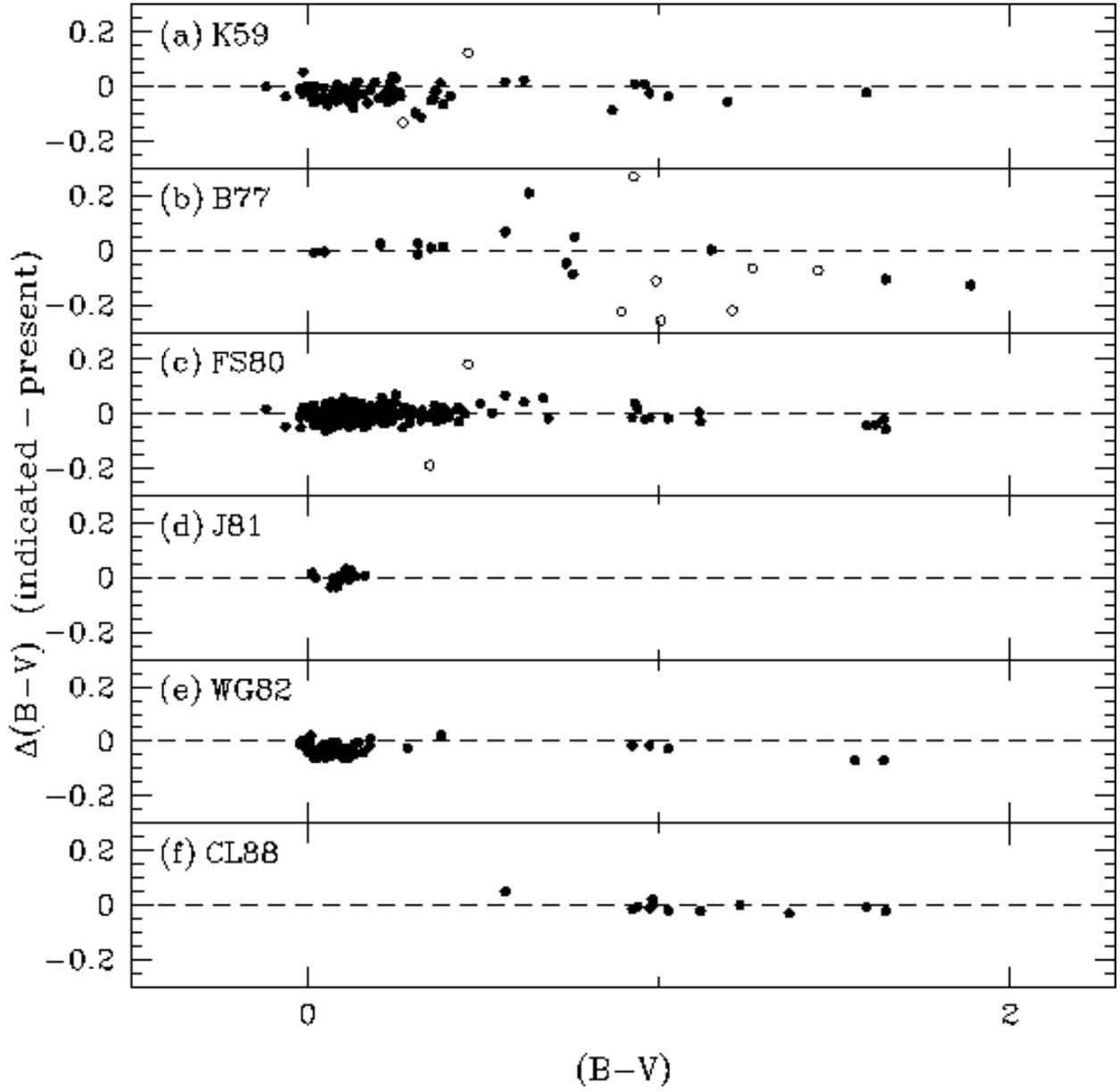}
\caption{Same as Figure \ref{fig:mkcompmag} except comparing $(B-V)$ colors.}
\label{fig:mkcompcolor}
\end{figure}

\begin{figure}
\plotone{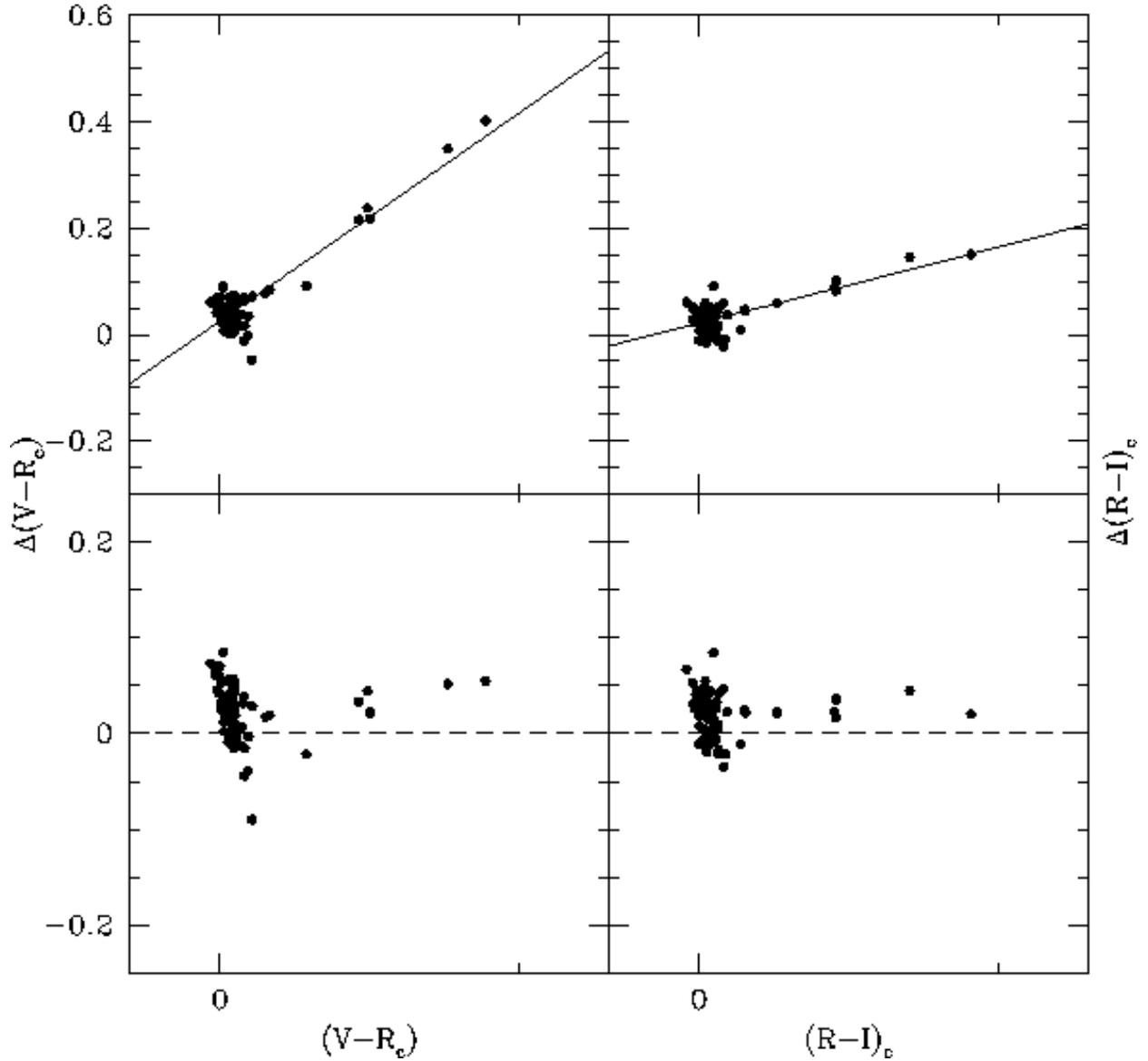}
\caption{Comparisons of our $(V-R_{c})$ and $(R-I)_{c}$ photometry with 
that given by \citet{{WizinowichGarrison1982}}.  The top panels plot 
the raw differences and show that strong systematics exist between the 
two data sets.  The solid lines denote the least-squares fit to the 
data.  The bottom panels are the differences that result once these 
systematics are removed.}
\label{fig:mkcompWG}
\end{figure}

The fact that NGC~3532 is projected against a very rich population of disk 
stars at low galactic latitude is clearly evident by the large amount of 
field star contamination in the CMDs shown in Figure \ref{fig:cmds_init}. 
As a result, we are forced to explore various techniques that would help 
to reduce the impact of this field population on the CMDs and to better 
define the cluster main sequence towards fainter magnitudes.

Stellar proper motions can be used as a robust and reliable tool to help 
determine cluster membership.  To this end, we have cross-referenced our 
photometry list with the stars in the UCAC3 catalog in an effort to better 
separate the field and cluster populations.  There are 18,131 stars in 
common, with 17,076 of these judged to be legitimate single stars as based 
on their object classification and double star flags in UCAC3 (see 
Zacharias et al. 2010\nocite{Zacharias2010} for details).  We further 
reduce the sample by 704 more stars that either have blank or null entries for 
their proper motions.

\begin{figure}
\plotone{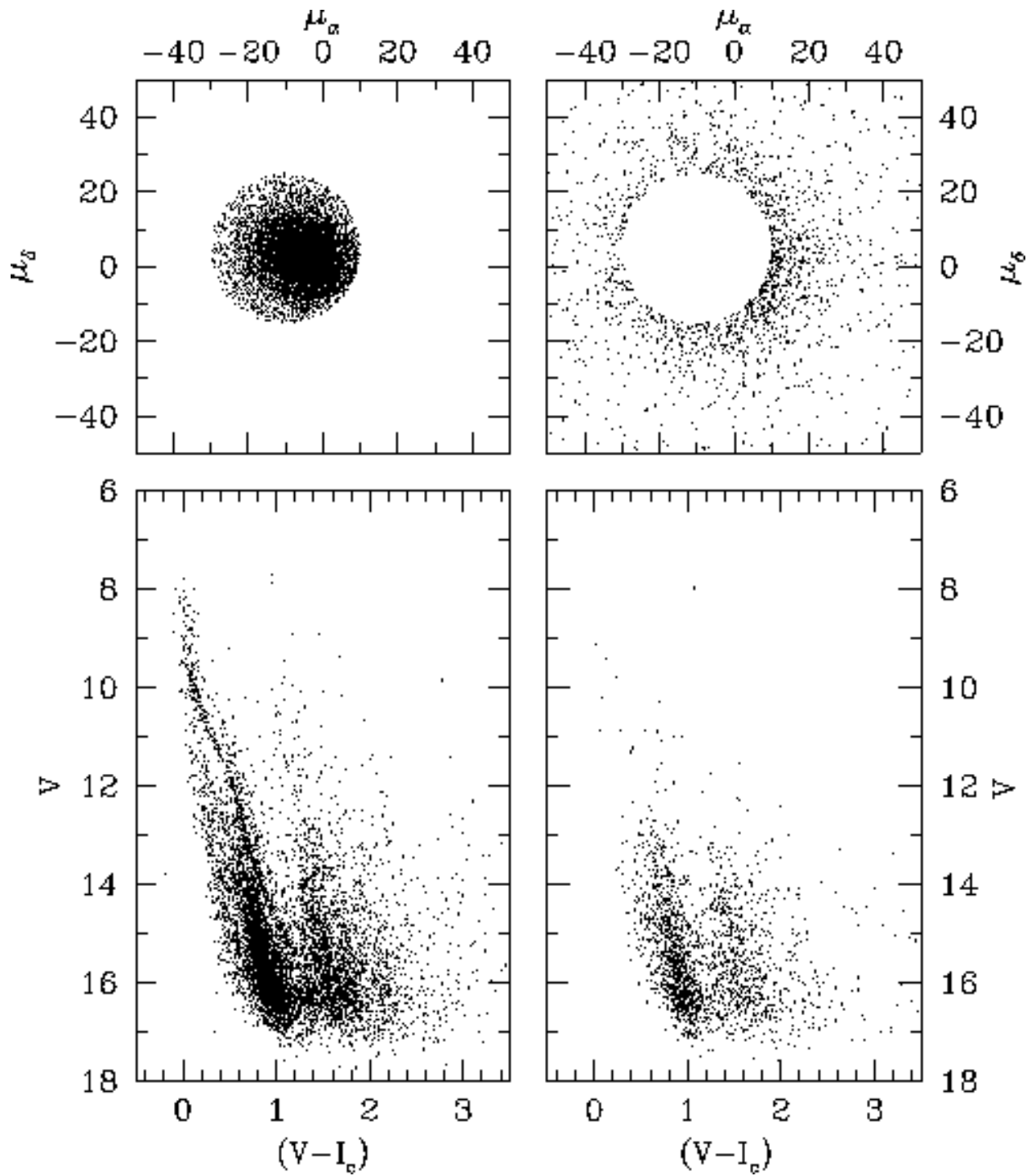}
\caption{Proper motion diagrams and corresponding $[V,~(V-I_{c})]$ CMDs for 
stars in common between our photometric data set and the UCAC3 catalog in 
the NGC~3532 field.  Probable cluster members are shown in the left-hand 
panels by plotting stars that are within 25~mas/yr of the cluster's mean 
proper motion, whereas the right-hand panels plot stars outside this radius.  
Note that while a number of stars not affiliated with NGC~3532 have been 
removed using this cut, a large field star population is still quite 
evident in the CMD in the lower-left panel.}
\label{fig:pmplot}
\end{figure}

In Figure \ref{fig:pmplot} we present the proper motion diagram that 
results from the UCAC3 data together with the $[V,~(V-I_{c})]$ CMD for 
stars in the NGC~3532 field that have complementary photometry.  In the 
left-hand panels we plot only those stars within a radius of 25 
mas~yr$^{-1}$ of the absolute proper motion for the cluster 
\citep[$\mu_{\alpha}=-10.04$, $\mu_{\delta}=+4.75$;][]{vanLeeuwen2009}, 
whereas the right-hand panels show stars outside this radius.  Although 
this selection criterion does succeed in removing some of the field star 
contamination, especially towards the faint end of the CMD, the fact that 
the cluster's absolute proper motion does not differ appreciably from 
that of the background field stars (assumed to be 
$\mu_{\alpha}\approx\mu_{\delta}\approx0$) poses some difficulty for our 
attempts to isolate the main sequence.  While a radius of 25 mas 
yr$^{-1}$ may seem to be a bit too generous, additional plots such as 
these, which are not shown here, that use selection radii of 10, 15, and 
20 mas yr$^{-1}$ did further succeed in reducing the number of field 
stars, but at the cost of excluding more and more stars that appeared to 
lie on or near the NGC~3532 main sequence, particularly towards the 
fainter magnitudes where errors in UCAC3 proper motions begin to become 
quite large.  Moreover, the magnitude limit of the UCAC3 catalog 
($V\sim17$) clearly does not extend to the faintest areas of the CMD 
where the field star population begins to totally obscure the cluster 
main sequence (c.f., Figure \ref{fig:cmds_init}).  In the end, we are 
resigned to accept that we cannot rely solely upon the UCAC3 proper 
motion information to procure a decent sample of bona fide cluster 
members, at least towards the fainter end of the CMDs where the field 
star contamination is the heaviest.

\begin{figure}
\plotone{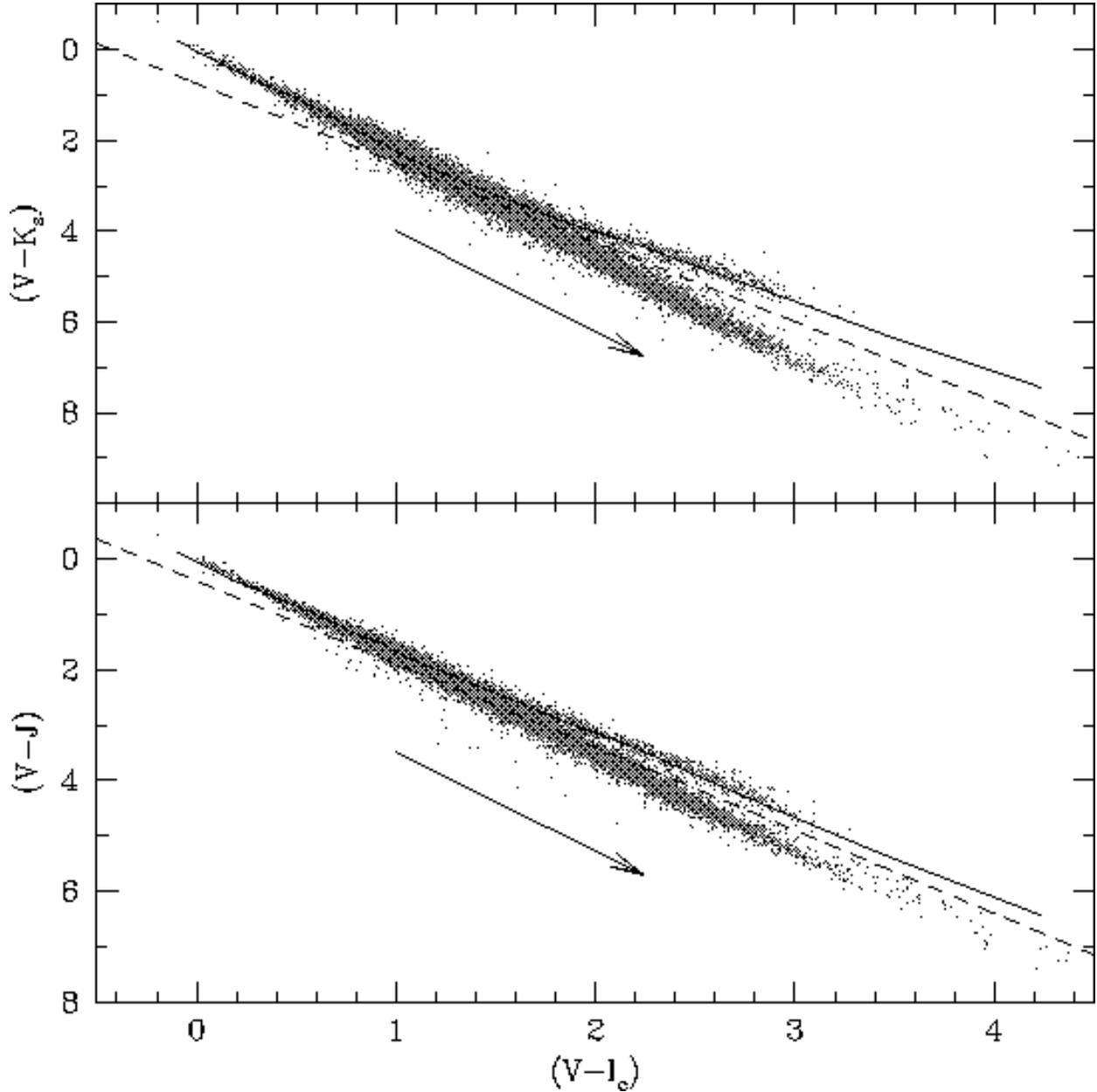}
\caption{Color-color diagrams on the $[(V-J),~(V-I_{c})]$ and 
$[(V-K_{s}),~(V-I_{c})]$ planes that result from a combination of our 
$BV(RI)_{c}$ photometry with $JHK_{s}$ data from the 2MASS catalog.  
Solid lines denote the standard relations, appropriate for dwarf stars.  
Based on the slopes of the reddening vectors, indicated by arrows in 
both panels, stars lying below the dashed lines correspond primarily to 
a population of highly-reddened field stars.}
\label{fig:ccremoval}
\end{figure}

A second attempt to reduce the field star contamination involved 
isolating the cluster sequence from the background stars by 
supplementing our data with infrared photometry.  To this end, we 
extracted the $JHK_{s}$ magnitudes and associated uncertainties for 
67,670 stars in common between our data set and the 2MASS catalog.  Our 
goal was to ascertain which combination of optical and/or infrared 
color indices produced the best possible separation between the field 
and cluster populations.  The results of this exercise are shown in 
Figure \ref{fig:ccremoval} where we have plotted two color-color 
diagrams for stars in the NGC~3532 field.  The $[(V-J),~(V-I_{c})]$ diagram 
in the bottom panel reveals a noticeable separation between two 
distinct sequences of stars toward the redward end of the plot.  
Likewise, the $[(V-K_{s}),~(V-I_{c})]$ diagram in the top panel shows this same 
separation, but to a somewhat greater degree.  To help us better 
identify which of these sequences actually belong to the cluster itself 
we have overplotted the standard relations for dwarf stars (solid 
lines) as given by \citet{BessellBrett1988}, transformed to the 2MASS 
system using the relations of \citet{Carpenter2001}, in both panels.  
Based not only on the reddening vectors indicated in the plots, but 
also the loci of dwarfs predicted by the standard relations, we 
conclude that the lower branch of stars lying at $(V-I_{c})\gtrsim1.5$ in 
both panels correspond primarily to the population of field disk stars 
with various reddenings.  The stars in the upper branch, on the other 
hand, include both legitimate cluster stars as well as some of the 
field dwarf stars that are situated on the blueward side of the main 
sequence in Figure \ref{fig:cmds_init}.

\begin{figure}
\plotone{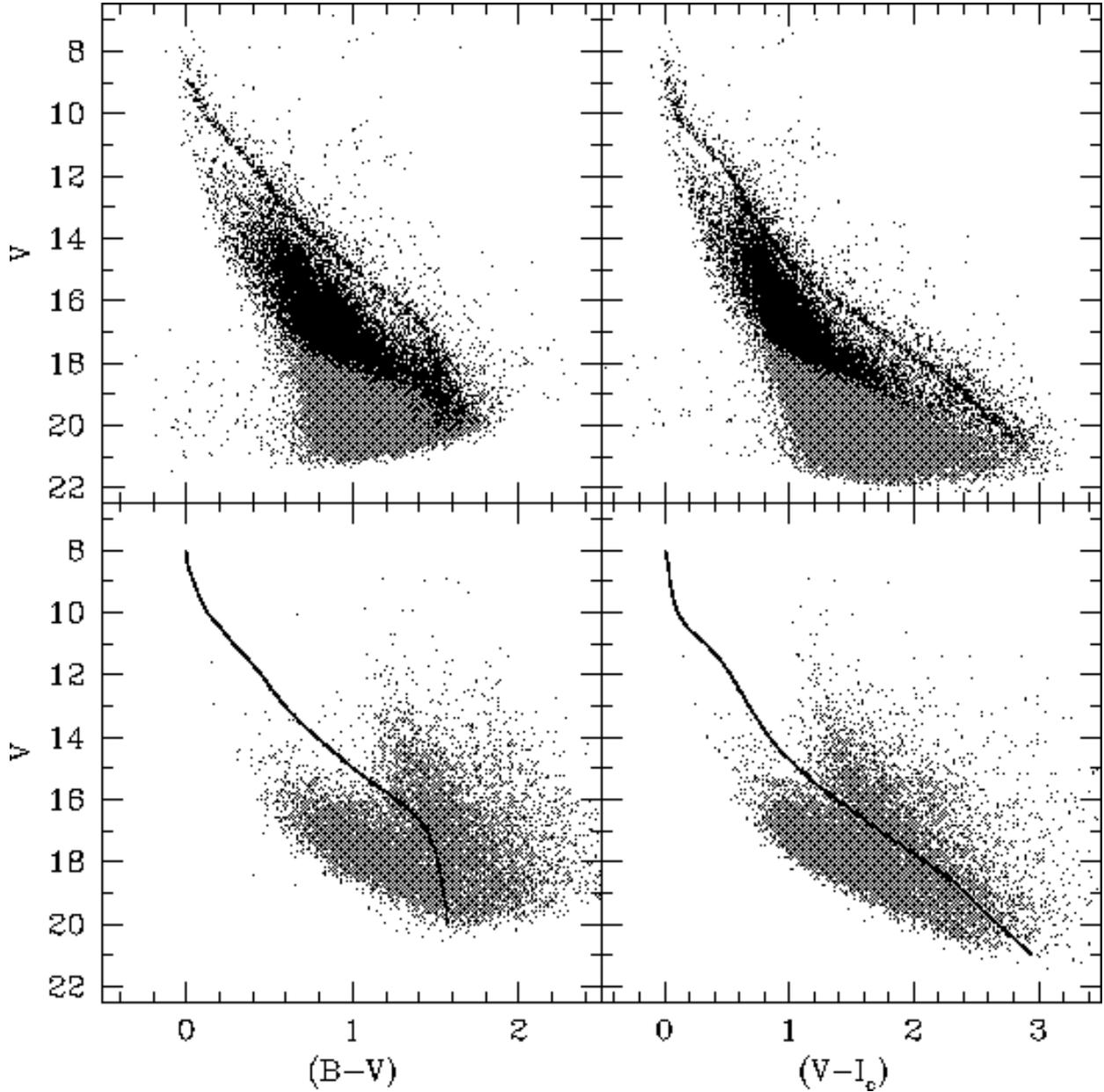}
\caption{CMDs for stars in the the NGC~3532 field once the cuts from 
Figure \ref{fig:ccremoval} are applied (top panels).  While gray dots 
represent stars having $BV(RI)_{c}$ photometry in our data set, black dots 
correspond to those that have complementary $JHK_{s}$ photometry from 2MASS.  
The bottom panels show stars that were removed using the cuts together 
with our derived main sequence fiducial for NGC~3532.}
\label{fig:cmds_cleaned}
\end{figure}

If we exclude stars lying below the dashed lines in Figure 
\ref{fig:ccremoval} and replot the CMDs for NGC~3532, as shown in the 
top row of Figure \ref{fig:cmds_cleaned}, then the cluster's lower main 
sequence stands out quite clearly against the remaining field star 
population and can easily be traced down to $V\sim20$.  It is important 
to note that we have used the same criteria for plotting the CMDs as in 
Figure \ref{fig:cmds_init} but with the addition of excluding stars 
that have uncertainties greater than 0.2 mag in their $JHK_{s}$ 
magnitudes.  Stars that satisfy these criteria are denoted by black 
dots in Figure \ref{fig:cmds_cleaned}, while those that have large 
uncertainties in $JHK_{s}$, or lack 2MASS photometry entirely, are 
shown as gray dots.  Note that the lack of field stars having $JHK_{s}$ 
photometry towards bluer colors in each CMD is due to the fact that 
these stars are just too faint to be detected by the 2MASS survey.  
Fortunately for us, however, their absence does not inhibit our 
investigation of the NGC~3532 main sequence since these stars primarily 
belong to the field population.

While our arbitrary selection of photometry based on the locations of 
stars in the color-color diagrams in Figure \ref{fig:ccremoval} has 
helped us to better define the cluster main sequence, the question remains 
as to whether our culling process may have also excluded some legitimate 
cluster members.  For this reason, we show in the bottom panels of Figure 
\ref{fig:cmds_cleaned} the stars that were rejected based on their 
locations below the lines in Figure \ref{fig:ccremoval}.  Unfortunately, 
the density of field stars in the vicinity of the lower main sequence is 
still too high to judge if any stars belonging to the cluster still 
remain, but based on the $BV(RI)_{c}$ photometry we have derived for the 
NGC~3532 field, in combination with the 2MASS $JHK_{s}$ magnitudes, this is 
likely the best method, given the information currently available, for 
separating the cluster sequence from the field stars at fainter 
magnitudes.

In order to benefit our analysis of NGC~3532 in later sections, we will 
further endeavor to isolate the cluster's main sequence from the field 
star population using a technique of photometric filtering.  Our aim is 
to identify a sample comprised of stars that have a high probability of 
belonging to NGC~3532 and is based on the fact that they should remain 
within a common ``envelope" in any given CMD, while the field stars 
will be scattered in and out of these envelopes depending on their 
reddenings and/or photometric uncertainties.

A first step in this process involves identifying the location of the 
main sequence in different CMDs by eye and removing stars that are more 
than 0.5 mag in color from this initial fiducial.  In subsequent 
iterations the fiducial colors are redetermined by taking the median 
value over a small range in $V$ magnitude and excluding objects lying 
more than $3\sigma$ in color away from this median.  The final fiducial 
is produced when these median values do not change appreciably from one 
iteration to the next, and the number of stars lying within the 
envelope defined by the fiducial remains constant.

Next, with a trial fiducial, we then compute a $\chi^{2}$ value for 
each object in a CMD as the following:

\begin{equation}
\chi^{2}=\sum\limits_{i=1}^N \frac{(\Delta\,color_{i})^{2}}{\sigma(color_{i})^{2}+\gamma_{i}\sigma(mag_{i})^{2}+\sigma_{0}^{2}},
\end{equation}

\noindent{where $\Delta\,color_{i}$ is the difference in color between the 
$i$th data point and the fiducial at the magnitude of the star, and 
$\sigma(color_{i})$ and $\sigma(mag_{i})$ are the photometric errors in 
color and magnitude, respectively.  The $\gamma_{i}$ term corresponds to 
the slope of the main sequence at the data point and is included to 
account for the error in color caused by the uncertainty in magnitude.  
Finally, the $\sigma_{0}$ term is included and adjusted to force the total 
$\chi^{2}$ to be roughly equal to the number of stars, N.}

Using this definition, we compute three separate $\chi^{2}$ values for 
each star using its $(B-V)$, $(V-I_{c})$ and $(V-K_{s})$ colors and $V$ 
magnitude.  We then reject stars if any of their $\chi^{2}$ values is 
greater than $3\sigma$.  Transversely, a given star will be tagged as a 
member of the cluster by this technique only if all three of its 
$\chi^{2}$ values determined from its $(B-V)$, $(V-I_{c})$, and $(V-K_{s})$ 
colors are within $3\sigma$.

\begin{figure}
\plotone{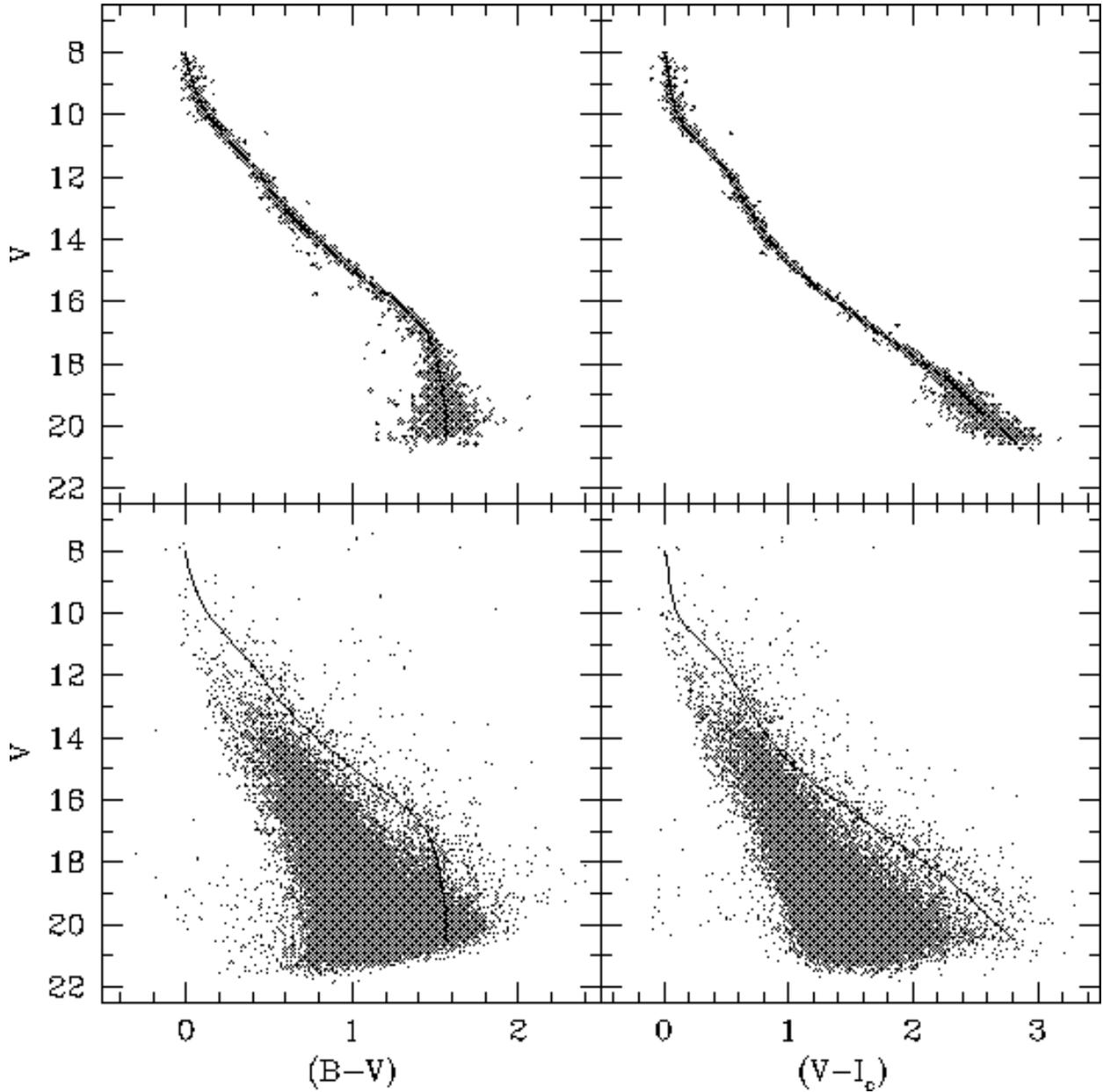}
\caption{CMDs for stars that have been identified as probable main 
sequence members of NGC~3532 using our photometric filtering technique 
(top panels).  Bottom panels show objects that have been rejected by the 
filtering algorithm as belonging primarily to the field.  While the 
bottoms panels reveal virtually no presence of main sequence stars in the 
vicinity of the fiducials (solid lines) that were inadvertently removed 
during this filter process, the presence of a binary star population 
belonging to NGC~3532 is still evident as a sequence of stars lying 
roughly parallel to the fiducials.}
\label{fig:mkgood1}
\end{figure}

The top row of Figure \ref{fig:mkgood1} presents CMDs for the final 
sample of main sequence stars that have been isolated from the field 
star population using our technique.  Alternatively, the bottom row 
shows only the rejected stars (i.e., the field stars).  Although a few 
obvious outliers still exist in the upper panels, our filtering has 
successfully removed a large number of field stars from the cluster 
main sequence.  An important note, however, is that the bottom panels 
still show the presence of the cluster binary stars lying as much as 
0.75 mag above the fiducial.  Unfortunately, our filtering technique 
does not account for the presence of binaries, and our final sample of 
cluster members will, by design, include predominately single stars.

\begin{figure}
\plotone{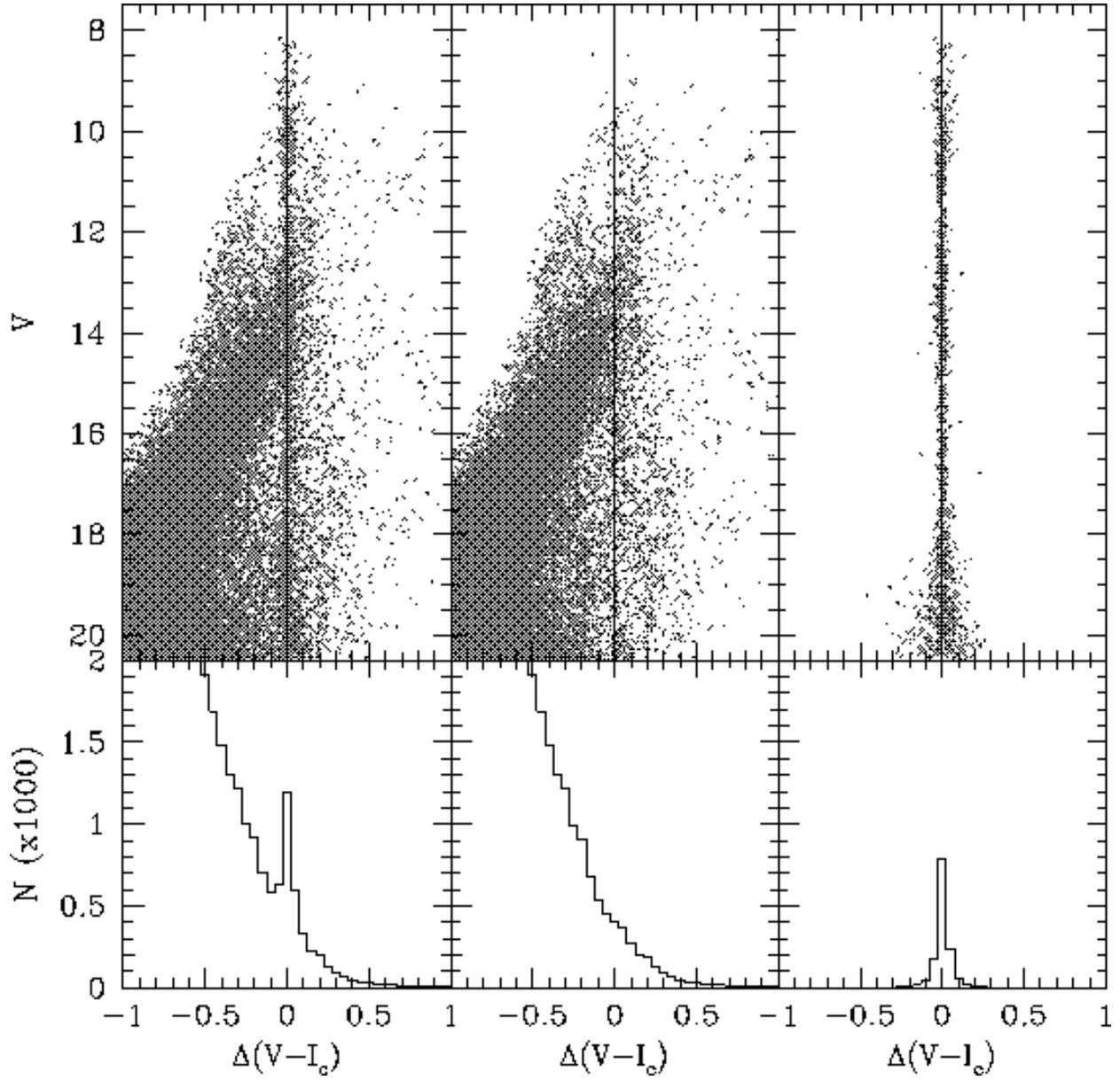}
\caption{$(V-I_{c})$ color difference between individual stars and our derived 
fiducial sequence for NGC~3532 along with corresponding histograms of 
their distributions (top and bottom panels, respectively).  From 
left-to-right the panels show all stars in the NGC~3532 field, stars 
removed by our photometric filtering technique, and stars identified as 
probable main sequence members of the cluster.}
\label{fig:mkgood2}
\end{figure}

An alternative means by which to test the robustness of our photometric 
filtering technique is to examine the distribution of stars as a function 
of $\Delta\,color$.  Figure \ref{fig:mkgood2} shows a few such plots that 
use $\Delta\,(V-I_{c})$ as the color of choice.  The top row of panels show, 
as a function of $V$ magnitude, the $\Delta\,(V-I_{c})$ values for all the 
stars, only the field stars, or only the cluster stars from left to right, 
respectively.  The bottom row reveals complementary histograms of these 
same distributions to illustrate that once the cluster stars are removed 
using our filtering process, the underlying field star population shows a 
fairly smooth transition over the cluster main sequence region from 
$-0.3\leq\Delta\,(V-I_{c})\leq0.3$.

\section{Discussion}
\label{sec:discussion}

\subsection{Previous Cluster Parameter Estimates}
\label{subsec:previousparams}

One of the main goals in our investigation of NGC~3532 is to derive new 
estimates for the cluster parameters (e.g., distance, reddening, age, 
etc.) that are based on the photometry presented herein.  Indeed, the 
depth and precision of our data compared to previous works offers a 
number of obvious advantages for our analysis.  Moreover, the 
combination of our $BV(RI)_{c}$ data with 2MASS $JHK_{s}$ photometry 
provides high-quality observations that span a wide wavelength range 
and allows us to utilize different color indices to better constrain 
values for the cluster parameters.

Before delving into our own determination of parameters for NGC~3532, 
however, it is helpful to first consider some of the values for the 
cluster's distance, reddening, age, and metallicity that have been 
derived by previous investigations.  Table \ref{tab:table6} presents a 
compilation of the various parameter estimates derived by the studies 
indicated in the first column.  The final column of the table gives a 
brief explanation of how these parameters were determined.  In some 
cases we also include the uncertainties in the values as quoted in the 
original investigations.

Interestingly, apart from the distances quoted by \citet{Robichon1999} 
and \citet{vanLeeuwen2009}, both of which were derived using $Hipparcos$ 
parallaxes, virtually all of the values for the distance moduli agree 
quite well with each other.  The one exception to this is the value of 
$(m-M)_0=8.06$ given by \citet{Johansson1981}, but this is likely a 
direct consequence of their higher reddening compared to the other 
studies listed.  If we instead assume $E(B-V)\sim0.04$, which better 
corresponds to the other reddening values listed in the third column, 
their distance modulus would increase to $(m-M)_0=8.26$ (assuming a 
$R_V=3.1$ reddening law).  The range in distance moduli tabulated in 
Table \ref{tab:table6} place the cluster between 400 and 500 pc from the 
Sun, making NGC~3532 one of the closest known open clusters.

The reddening for NGC~3532 likewise seems be well constrained with 
$E(B-V)$ estimates from the various studies ranging between 0.01 and 
0.1.  Note that we have converted the $E(b-y)$ values given by 
\citet{Eggen1981}, \citet{Schneider1987}, and \citet{Malysheva1997} to 
$E(B-V)$ assuming $E(B-V)=1.35E(b-y)$ \citep{Crawford1975}.  Despite its 
position very near the Galactic plane, the low reddening for NGC~3532 is 
consistent with its proximity to the Sun.  For comparison, the reddening 
maps of \citet{Schlegel1998} give a full line of sight reddening of 
$E(B-V)\sim1.20$ at Galactic coordinates corresponding to the center of 
the cluster.  Multiplying this value by a factor of 
$(1-e^{-|d\,sin\,b|/h})$, where $d$ is the cluster's distance, $b$ is 
its Galactic latitude, and $h$ the scale height of the typical dust 
layer \citep[assumed to be 125 pc;][]{Bonifacio2000} results in 
$E(B-V)\sim0.1$ if $d=450$ pc.  While this is a somewhat crude upper 
limit to the reddening for NGC~3532, it serves to support the remarkably 
low $E(B-V)$ values that have been previously suggested.

Four independent estimates for the cluster's metallicity exist in the 
literature.  While two of these have been derived from high-resolution 
spectroscopic analysis of cluster giant stars \citep{Luck1994,
  Gratton2000}, the other two given by \citet{Piatti1995} and 
\citet{Twarog1997} are based on calibrations of DDO color indices 
versus [Fe/H].  Reassuringly, all four [Fe/H] determinations are 
in good agreement and point to a near-solar metallicity for the 
cluster. For this reason, we will adopt [Fe/H]=$0.0\pm0.1$ for 
NGC~3532 in the subsequent analysis.

\subsection{Empirically Derived Distance and Reddening}
\label{subsec:params}

We begin our determination of the distance and reddening for NGC~3532 
by comparing the locus of cluster stars in the $(B-V)$, $(V-I_{c})$, and 
$(V-K_{s})$ CMDs to the well-established main sequence for the Hyades.  The 
Hyades sample used here is based on the collection of so-called 
``hi-fidelity'' members presented by \citet{deBruijne2001}, who derived 
distances to individual stars based on secular parallaxes.  These 
secular parallaxes are $\sim2-3$ times more precise than the original 
parallax estimates given in the $Hipparcos$ catalog, thus providing 
better constraints on the absolute magnitudes of the Hyades members.  
The photometry we employ for the Hyades comes from 
\citet{Pinsonneault2004} who assembled $BVI_{c}JHK_{s}$ data from 
various sources to produce a catalog of color indices for the 
\citet{deBruijne2001} ``hi-fi" sample.  There is, however, one point to 
make regarding the photometry tabulated by \citet{Pinsonneault2004}, in 
that \citet{TaylorJoner2005} found systematic differences between their 
$(V-I_{c})$ data and Pinsonneault et al.'s for Hyades stars in common.  
This result is strengthened when comparing the homogeneous $BV(RI)_{c}$ 
photometry for a large sample of Hyades stars given by 
\citet{Joner2006} to the photometry tabulated by 
\citet{Pinsonneault2004} (c.f., Fig 23 in An et al. 
2007\nocite{An2007}).  For this reason, we have opted to use the 
$(V-I_{c})$ photometry of \citet{Joner2006} for 42 out of 92 stars in the 
Hyades sample, while the $(V-I_{c})$ photometry for the remaining stars 
comes from \citet{Pinsonneault2004}, but transformed to the 
\citet{Joner2006} system using a simple linear relationship.

\begin{figure}
\plotone{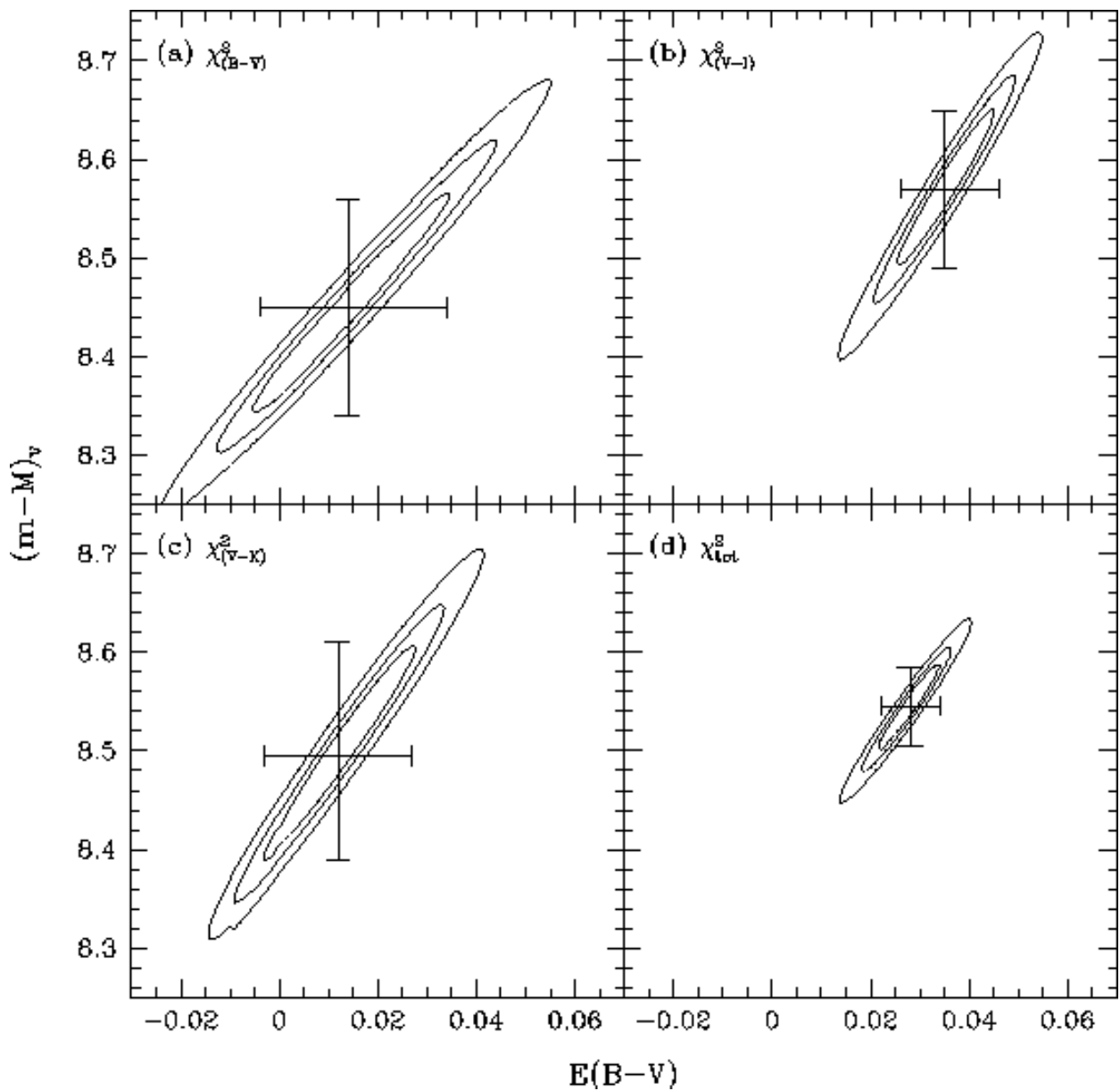}
\caption{$\chi^{2}$ contours resulting from our fits of the Hyades main 
sequence to the NGC~3532 fiducial to derive the cluster's distance and 
reddening.  The contours in panels (a), (b), and (c) correspond to fits 
using either the $(B-V)$, $(V-I_{c})$, and $(V-K_{s})$ colors, 
respectively, as ordinates in the CMDs, whereas panel (d) denotes the 
results by combining the $\chi^{2}$ values from all 3.  The horizontal 
and vertical errors bars represent approximate uncertainties in 
reddening and distance, respectively, when each are determined 
separately.}
\label{fig:contours}
\end{figure}

The distance and reddening for NGC~3532 are derived simultaneously by 
fitting the main sequence fiducial for the cluster, derived in Section 
\ref{subsec:ms}, to a sample of unevolved Hyades members (i.e., stars 
having $(B-V)>0.39$, $(V-I_{c})>0.44$, and $(V-K_{s})>0.98$).  This technique 
is akin to the one described by \citet{Richer1997} in their derivation 
of the distance and reddening to the globular cluster M~4.  To account 
for the metallicity difference between NGC~3532 ([Fe/H]=$0.0\pm0.1$) 
and the Hyades \citep[[Fe/H=$0.13\pm0.01$;][]{Paulson2003}, we adjust 
the colors of the cluster fiducial redward by $(B-V)=0.029$, 
$(V-I_{c})=0.015$, and $(V-K_{s})=0.027$.  The $\chi^2$ goodness of fit 
contours are shown in Figure \ref{fig:contours} from separately fitting 
our NGC~3532 fiducial to the empirical Hyades main sequence on the 
$[M_{V},~(B-V)_{0}]$, $[M_{V},~(V-I_{c})_{0}]$, and $[M_{V},~(V-K_{s})_{0}]$ 
planes (i.e., $\chi^2_{(B-V)}$, $\chi^2_{(V-I_{c})}$, and $\chi^2_{(V-K_{s})}$, 
respectively).  The lower right-hand panel of the same figure shows the 
combination of $\chi^2$ values resulting from all 3 fits 
($\chi^2_{tot}$).  The contours in each panel designate the 68.3\%, 
95.4\%, and 99.7\% (1, 2, and 3$\sigma$) confidence levels of our fits 
based on 2 free parameters while the error bars correspond to the 
uncertainties in distance and reddening when each parameter is 
considered separately.

Clearly, the combination of fits involving all 3 color indices yields 
much tighter constraints on the cluster distance and reddening.  The 
parameters that minimize the $\chi^2_{tot}$ distribution shown in the 
lower right-hand panel of Figure \ref{fig:contours} occur at 
$E(B-V)=0.028(\pm0.006)$ and $(m-M)_V=8.54(\pm0.04)$.  This translates 
to $E(V-I_{c})=0.039(\pm0.008)$, $E(V-K_{s})=0.083(\pm0.018)$, and 
$(m-M)_0=8.45(\pm0.05)$ when using the reddening coefficients for 
different bandpasses given by \citet{Schlegel1998} assuming a $R_V=3.1$ 
reddening law.  These values are in superb agreement with most of the 
modern distance and reddening estimates listed in Table 
\ref{tab:table6} for NGC~3532 that use photometry for their 
derivations.  Moreover, our estimated distance modulus is consistent 
with those derived from $Hipparcos$ parallaxes to within 2 sigma.

\subsection{Upper Main Sequence, Giant Stars, and Age}
\label{subsec:giants}

With the good constraints on the distance and reddening for NGC~3532, we 
can now use theoretical models to estimate the age of the cluster. 
Generally, this method involves determining which isochrone model of an 
appropriate metallicity best reproduces both the shape of the cluster 
turnoff as well as the luminosity of the red giant stars.  In practice, 
this method seems simple, but in reality the reliable modeling of main 
sequence and giant stars, particularly for a cluster seemingly as young 
as NGC~3532, is sometimes problematic due to various physics involved in 
computing the theoretical models for the different types of stars.

Our CMDs for NGC~3532 in Figure \ref{fig:cmds_init} show the cluster 
has a well populated turnoff region, but only a handful of giant stars.  
In addition, the fact that the cluster is situated in a region of high 
field star density means that some of the assumed red giant stars in 
the upper right portion of the CMDs may not belong to the cluster at 
all.  For this reason, we have isolated a sample of candidate red 
giants from the cluster CMDs for further analysis.  Assuming the 
cluster giants have $(B-V)>0.25$ and $V<8.5$, we find that 14 stars 
from our database meet these criteria.  The photometric, astrometric, 
and kinematic properties of these 14 candidate giant stars are listed 
in Table \ref{tab:table7}.  The information presented in this table is 
meant to include or exclude these stars as bona fide members of the 
cluster based on their observed properties.  We also include $(U-B)$ 
photometry in the table from FS80.  In addition, the proper motion 
information was extracted from the Tycho-2 catalog \citep{Hog2000} with 
radial velocities from \citet{GonzalezLapasset2002} and 
\citet{Mermilliod2008}.

\begin{figure}
\plotone{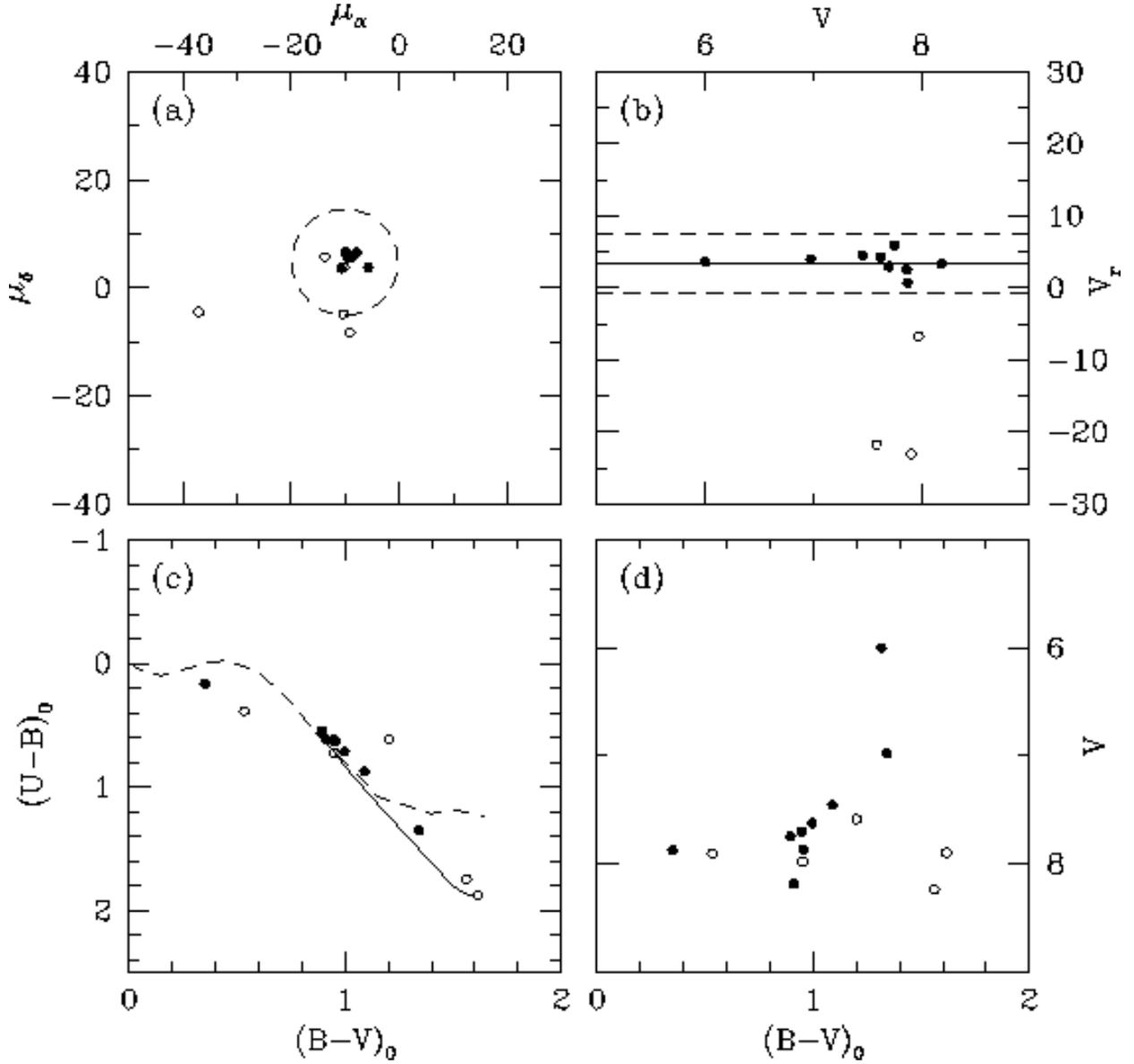}
\caption{Plots illustrating the photometric and kinematic properties of 
14 stars lying in the giant star region of the NGC~3532 CMDs.  In all 
panels we have plotted stars that have a high probability of belonging 
to NGC~3532 as solid circles, while those that likely belong to the 
field as open circles.  Panel (a) shows the proper motion 
characteristics of these stars using data taken from the Tycho-2 
catalog, with the dashed circle corresponding to a radius of 10~mas/yr.  
Panel (b) gives the corresponding radial velocity values of these stars 
as a function of $V$ magnitude, with the mean cluster velocity denoted 
as a solid horizontal line.  Stars within $\pm 4~$km/s of this mean 
(dashed lines) are likely cluster members.  Panels (c) and (d) show the 
$[(U-B),~(B-V)]$ and $[V,~(B-V)]$ diagrams, respectively, for these 
stars and are meant to illustrate that our cuts based on the kinematic 
properties have removed a number of objects that likely belong to the 
field.  The standard relations for dwarf and giant stars are shown as 
dashed and solid lines, respectively, in panel (c).}
\label{fig:giants} 
\end{figure}

In Figure \ref{fig:giants} we provide several different plots 
illustrating our membership selection criteria.  In the top two panels 
we use the kinematic information from Table \ref{tab:table7} to isolate 
members from non-members.  Specifically, we consider a star to be a 
member of NGC~3532 if its proper motion is within a radius of 10 mas 
yr$^{-1}$ from the cluster mean 
$(\mu_{\alpha},~\mu_{\delta})=(-10.04\pm0.24,~+4.75\pm0.21)$ 
\citep{vanLeeuwen2009} and it has a radial velocity within $\pm4$ km 
s$^{-1}$ of the mean of $V_{r}=+3.4$ km s$^{-1}$ 
\citep{GonzalezLapasset2002}.  Based on the $[(U-B),~(B-V)]$ 
color-color diagram and $[V,~(B-V)]$ CMD in the lower panels of Figure 
\ref{fig:giants}, it would appear that the stars we have excluded based 
on their kinematic properties alone correspond mostly to those that lie 
far from the primary cluster sequences.  As a result of this exercise, 
we conclude that 9 of the 14 stars in the giant region of the CMD are 
actual cluster members.

With a sample of giant stars that are legitimate members of the cluster, 
we can now move to fitting model isochrones to the photometry for NGC~3532 
to derive its age.  For this purpose, we employ the latest BaSTI stellar 
evolutionary models \citep{Pietrinferni2004} for two reasons.  First, they 
are among the most current available in terms of input physics and 
color-temperature relations and offer models that include convective core 
overshooting.  Secondly, the BaSTI models treat all evolutionary phases, 
including white dwarf cooling sequences \citep{Salaris2010}; this fact 
proves advantageous in the next section when we will compare such models 
to the observed properties of white dwarfs in NGC~3532.

\begin{figure}
\plotone{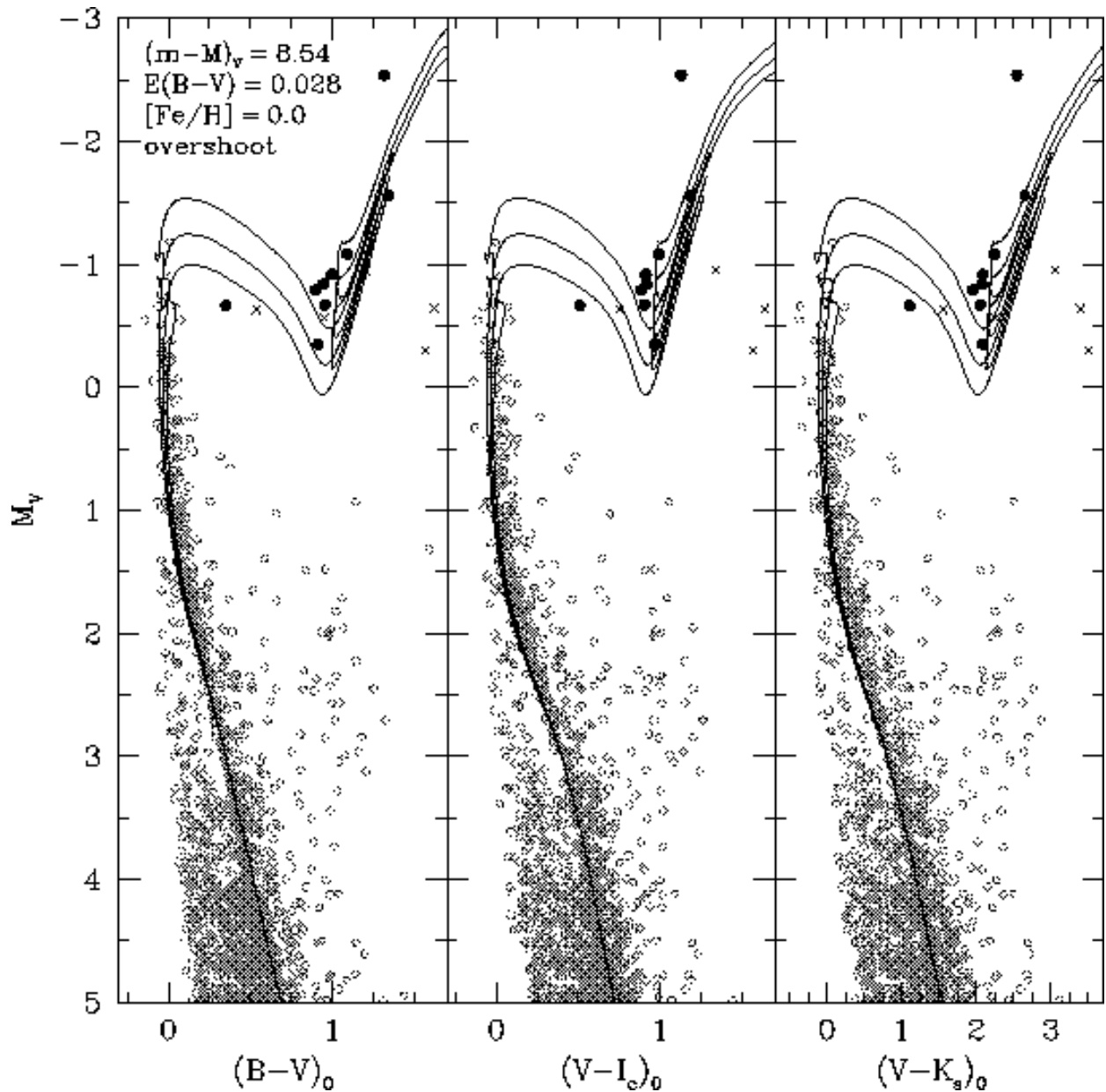}
\caption{Fits of solar-metallicity BaSTI isochrones to upper main 
sequence, turnoff, and giant star regions of the NGC~3532 CMDs.  The open 
circles are stars that belong either to the cluster main sequence or to 
the field, solid black circles represent objects that correspond to the 
cluster's giant star population based on Figure \ref{fig:giants}, and 
black crosses are stars that are likely non-members from the same figure.  
The fits here employ overshooting isochrones with an ages of 250, 300, and 
350\,Myr and use the distance and reddening derived from fitting the 
cluster's main sequence to the Hyades.  The agreement for stars near the 
turnoff is quite good, but the overshooting isochrone predicts a giant 
branch that is slightly too red when compared to the giant stars in 
NGC~3532.}
\label{fig:fit_over}
\end{figure}

\begin{figure} 
\plotone{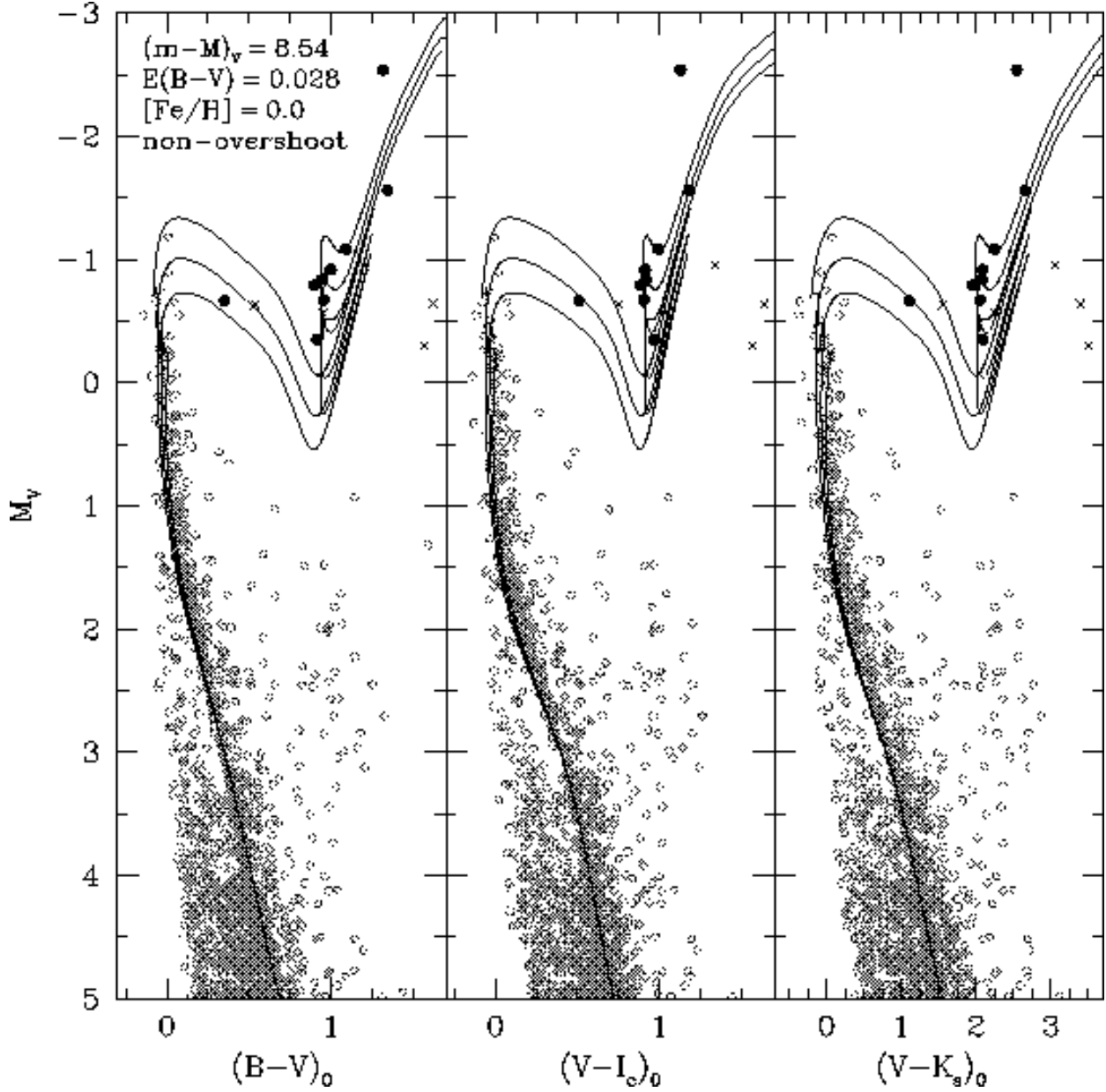} 
\caption{Same as Figure \ref{fig:fit_over}, but using non-overshooting 
isochrones from the BaSTI library.  These isochrones, which have ages of 
200, 250, and 300\,Myr, arguably do a better job of fitting the cluster 
giant stars, but the fits to the turnoff region and upper main sequence 
are less satisfactory than in Figure \ref{fig:fit_over}.}
\label{fig:fit_noover} 
\end{figure}

Based on the results presented in previous sections, we have chosen to 
fix the parameters for NGC~3532 at $(m-M)_{V}=8.54$, $E(B-V)=0.028$, and 
[Fe/H]=0.0 and explore which isochrones from the BaSTI library provide 
the best fit to the turnoff.  In Figures \ref{fig:fit_over} and 
\ref{fig:fit_noover} we compare such isochrone models, both with and 
without treatment of convective core overshooting, to the observed CMDs 
for the cluster.  For the $(V-K_{s})$ CMD we have transformed the 
$(V-K)_{J}$ colors of the BaSTI isochrones to the 2MASS system using the 
relations of \citet{Carpenter2001}.  The overshooting models with ages of 
250, 300, and 350\,Myr shown in Figure \ref{fig:fit_over} appear to fit 
the turnoff and upper main sequence of the cluster quite nicely, but they 
tend to lie systematically redward of the giant stars (filled circles).  
The non-overshooting models in Figure \ref{fig:fit_noover} with ages of 
200, 250, and 250\,Myr, on the other hand, arguably do a better job of 
matching the colors of the giant stars, but they lie consistently on the 
blueward side of the main sequence for $M_V\leq2.5$ in all three CMDs.  
Moreover, although not shown here, we have employed the isochrones of 
\citet{Girardi2002} to fit the observed CMDs and discovered that they 
provide virtually the same interpretation of the data as the BaSTI 
models.

Given that the location and shape of the giant branch is greatly 
influenced by the adopted convective mixing length in the model 
computations and color transformations, we are inclined to favor a 
cluster age that is derived mainly from fitting the upper main sequence 
and turnoff.  Therefore, since the overshooting isochrone provides a 
better fit to the CMDs in these regions, we estimate the age of NGC~3532 
to be $\sim300\,$Myr.

\subsection{White Dwarfs} 
\label{subsec:wds}

The existence of white dwarfs in NGC~3532 has been known for quite some 
time.  In their systematic search for white dwarf stars in young open 
clusters, \citet{ReimersKoester1989} used deep photographic observations 
of the cluster to identify 7 objects as candidate white dwarfs.  
Subsequent spectroscopic followup of these objects revealed 3 of them to 
be legitimate white dwarfs belonging to the cluster.  Spurred by the 
surprising lack of cluster white dwarfs in their initial study, 
\citet{KoesterReimers1993} revisited their search by enlarging the 
observed area around NGC~3532.  Indeed, they identified 3 additional 
objects as cluster white dwarfs, thus bringing the total number to 6.

More recently, \citet{Dobbie2009} presented a thorough analysis of these 6 
known white dwarfs using low-resolution, high signal-to-noise 
spectroscopy.  Their observations allowed them to derive precise 
temperature and surface gravity estimates by fitting spectral models to 
the observed Balmer line profiles.  Moreover, their inclusion of $V$-band 
photometry of the cluster field, extending to $V\sim20.5$, permitted a 
reevaluation of membership and lead to the conclusion that only four of 
the six white dwarfs discovered by \citet{ReimersKoester1989} and 
\citet{KoesterReimers1993} have distance moduli comparable to the cluster 
itself.  Furthermore, \citet{Dobbie2010} has identified, via spectroscopy, 
3 more faint blue objects as possible white dwarfs that may belong to the 
cluster.

Given that our $[V,~(B-V)]$ CMD for NGC~3532 in Figure 
\ref{fig:cmds_init} shows an abundance of objects at the faint blue 
end, it is worthwhile investigating the possibility that a large number 
of these may be white dwarf members of the cluster.  Such an analysis 
will be beneficial for two reasons.  First, increasing the number of 
known white dwarf stars in open clusters places tighter constraints not 
only on the masses of their progenitor stars, but also on the 
initial-final mass relationship.  Secondly, the comparison of the 
location of these stars in the CMD with white dwarf cooling models 
allows us to derive a cluster age that is free from uncertainties in 
the treatment of convective core overshooting.  Age estimates such as 
this have only been possible recently with the use of wide field 
observations obtained on large aperture telescopes that are able to 
extend faint enough to reach the end of the white dwarf cooling 
sequence in nearby open clusters (see, for example, the works of 
Kalirai 2001a\nocite{Kalirai2001a}, 2001b\nocite{Kalirai2001b}).

\begin{figure}
\plotone{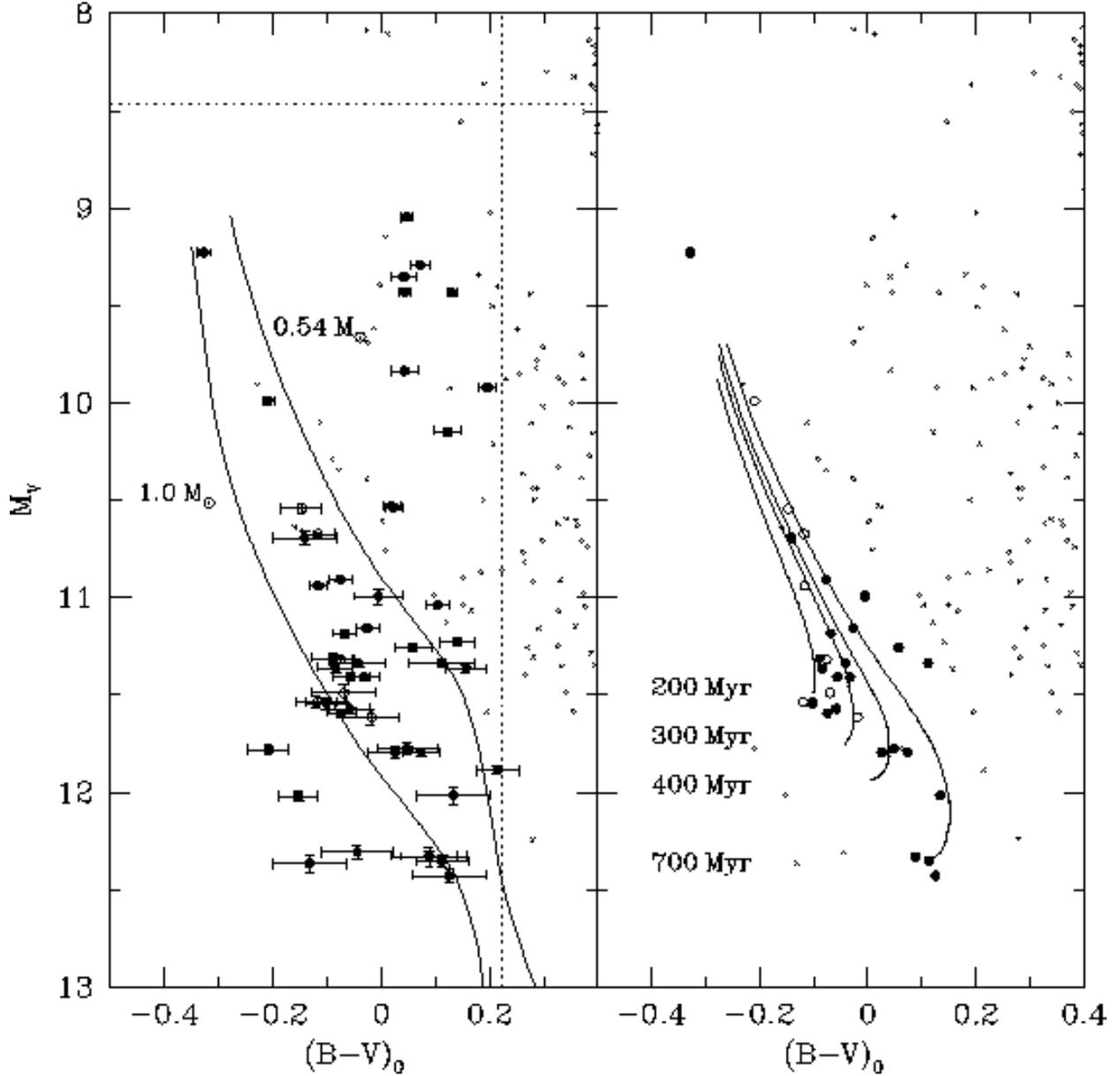}
\caption{Locations of a number of faint, blue objects in the NGC~3532 
field on a dereddened, distance-corrected CMD.  Dotted horizontal and 
vertical lines in the left-hand panel represent arbitrary cuts that were 
made to identify objects for further consideration.  Open and solid 
circles situated on the faint, blueward sides of these lines correspond 
to objects that were visually inspected in CCD images to ensure their 
stellarity.  The solid lines in the same panel are cooling models from the 
BaSTI evolutionary library for $0.54~M_{\odot}$ and $1.0~M_{\odot}$ white 
dwarfs.  Stars between these two tracks are probable white dwarfs belonging 
to NGC~3532 and are again represented in the right-hand panel as either 
open or filled circles.  Open circles correspond to the 8 previously 
identified white dwarfs, whereas filled circles are objects that are 
delineated in the left-hand panel.  BaSTI white dwarf isochrones having 
ages of 200, 300, 400, and 700\,Myr are also overplotted in this panel.} 
\label{fig:isolate} \end{figure}

In Figure \ref{fig:isolate} we present an enlarged portion of the cluster 
$[V,~(B-V)]$ CMD centered on the region where white dwarfs should reside.  
To single out potential cluster members, we begin by considering only those 
objects having $(B-V)\leq0.25$ and $V\geq17$.  As shown in the figure 
(denoted by dotted vertical and horizontal lines once reddening and 
distance have been accounted for), these limits exclude the vast majority 
of the field dwarf stars that contaminate the CMD below the cluster main 
sequence.  These criteria leave us with 78 objects that have photometry of 
reasonably high precision (i.e., assuming the limits of 
$\sigma(mag)\leq0.1$).  Upon visually inspecting these candidates in the 
cluster finding chart (c.f., Figure \ref{fig:fieldplot}), we can 
immediately exclude 28 of them as false detections due to their close 
proximity to much brighter stars, diffraction spikes, or known defects in 
the CCD.

The remaining 50 objects are shown in the left-hand panel of Figure 
\ref{fig:isolate} as either filled or open circles with error bars 
denoting their photometric uncertainties.  Note the prominent clump of 
objects lying at $(B-V)_{0}\sim-0.1$ and $M_{V}\sim11.5$ in the figure.  To 
help us further isolate more of these objects as potential cluster white 
dwarfs, we overplot the BaSTI cooling models for $M_{WD}=0.54\,M_{\odot}$ 
and $1.0\,M_{\odot}$ \citep{Salaris2010} in the figure and exclude stars 
that lie outside of the region they bracket.  These 32 remaining stars are 
shown in the right-hand panel of Figure \ref{fig:isolate} with filled 
circles denoting stars that have high probability of being new cluster 
white dwarfs while open circles shown the location of known white dwarfs 
in the cluster field from previous studies \citep{ReimersKoester1989, 
KoesterReimers1993, Dobbie2010}.  Also plotted in this panel are the BaSTI 
white dwarf isochrones for 200, 300, 400, and 700\,Myr.  Clearly, the 
three youngest isochrones terminate at absolute magnitudes comparable to 
the faintest members of the clump.  The 700\,Myr isochrone, on the other 
hand, does terminate at the location of the faintest candidates in our 
sample, but the bright end of the same isochrone extends too far to the 
red from the known white dwarfs in the cluster for us to accept such a 
high age.

Based upon our photometric uncertainties at the faint end (approaching 
$\sim0.1$ mag for stars as faint as $V\sim20$), together with the 
uncertainty in our derived cluster distance modulus, we can deduce that 
the white dwarf cooling age for the cluster is somewhere between 200 
and 400\,Myr.  Note that this age range corresponds nicely to the age 
derived when fitting the cluster turnoff when using either the 
overshooting or non-overshooting models.  However, the quality of our 
photometry, combined with the paucity of white dwarfs in the faint, 
blue region of the NGC~3532 CMD, prevents us from placing tighter 
constraints on the cluster's age using this technique.  Undoubtedly, it 
would prove useful to obtain high-quality spectroscopic observations of 
our white dwarf candidates to better locate the termination of the 
white dwarf cooling sequence.

\subsection{Binary Stars}
\label{subsec:binaries}

The impact of unresolved binary stars on a cluster CMD is commonly seen 
as the broadening of the main sequence towards brighter magnitudes.  
Indeed, it is well known that clusters containing a large number of 
equal-mass binary systems exhibit a secondary main sequence displaced by 
as much as $\sim-0.75$ mag relative to the single-star main sequence 
(see, for example, the CMDs for M~67 in Montgomery et al. 
1993\nocite{Montgomery1993}).  

\begin{figure}
\plotone{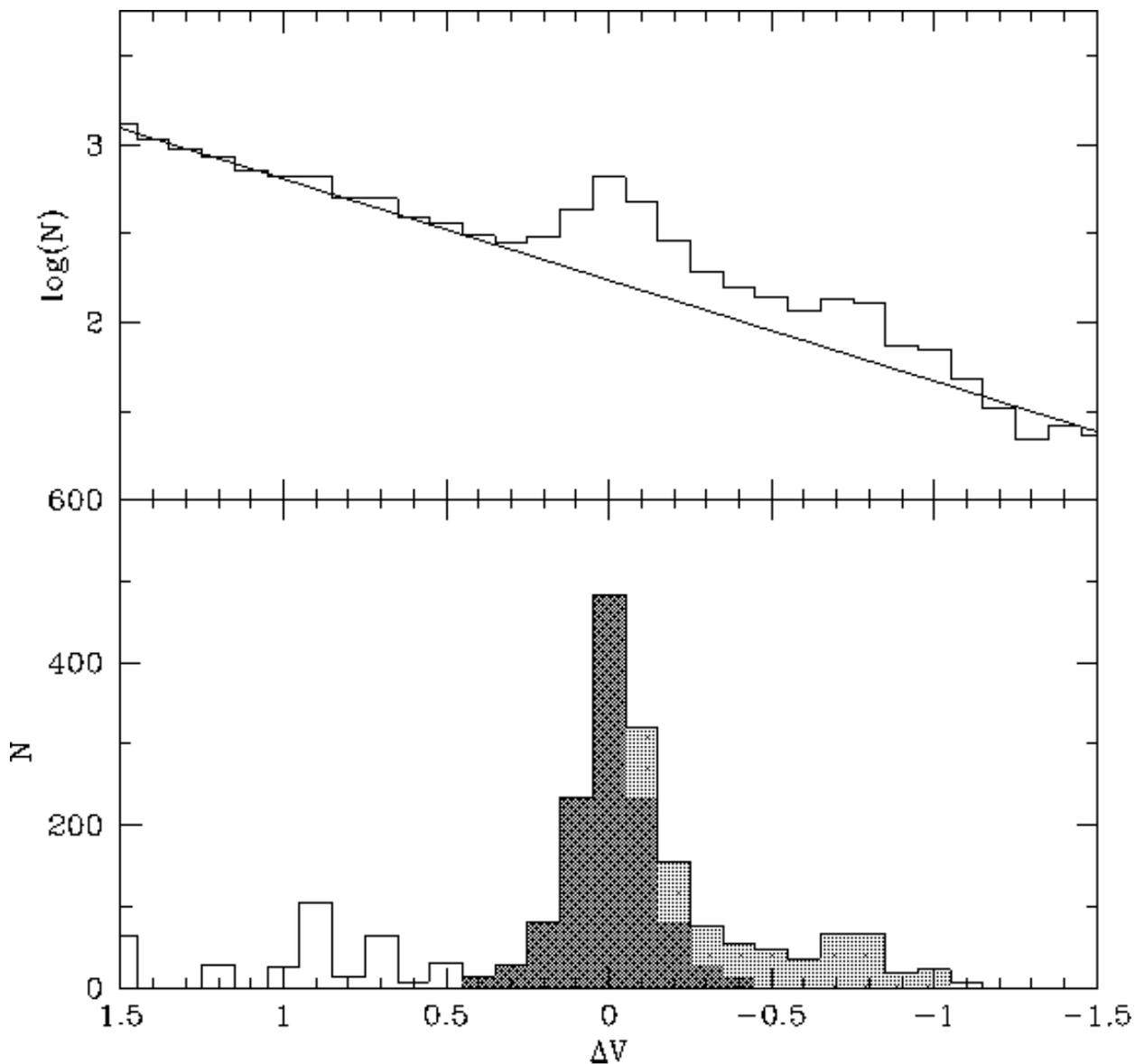}
\caption{Distribution of the number of objects as a function of 
$V$-magnitude difference relative to our derived fiducial sequence for 
NGC~3532 (top panel).  The straight line, meant to compensate for the 
field star population in the CMDs, has been fitted to this distribution in 
regions outside the envelope occupied by single and binary stars on the 
cluster's main sequence (i.e., $1.0<|\Delta V|<2.0$).  The 
result of this fit is shown in the bottom panel with the dark-gray shaded 
region corresponding to single stars lying on the main sequence, and the 
light-gray shaded area denotes the cluster's presumed binary star 
population.}
\label{fig:binaries}
\end{figure}

The existence of binary stars in NGC~3532 is most clearly evident in 
Figure \ref{fig:cmds_cleaned} as a scattering of dots lying parallel to 
the cluster main sequence in our CMDs.  To determine the approximate 
binary fraction for NGC~3532 we begin by taking the difference in $V$ 
magnitude ($\Delta\,V$) between our derived fiducial sequence and the 
individual stars within the range of $0.1\leq(V-I_{c})\leq2.9$.  The resulting 
histogram of the number of stars as a function of $\Delta\,V$ is shown in 
the upper panel of Figure \ref{fig:binaries}.  While the peak located at 
$\Delta\,V=0$ corresponds to the single stars in NGC~3532, the binary 
population easily reveals itself as the excess in the distribution towards 
brighter magnitudes with the secondary peak at $\Delta\,V\sim-0.7$ caused 
by a small number of equal mass systems.  Before we can derive a robust 
estimate for the binary fraction, however, we must compensate for the 
field star contribution to the histogram.  This has been done by fitting a 
linear relationship to the distribution of bins outside the area occupied 
by singles and binaries (specifically, $1.0<|\Delta\,V|<2.0$).  Once the 
field star contribution is removed, the resulting histogram, shown in the 
lower panel of Figure \ref{fig:binaries}, $should$ be a reasonable 
representation of both the single and binary star population in NGC~3532.

Assuming that the distribution of single stars approximates a Gaussian, 
we have simply reflected the bins at $0.0\leq\Delta\,V\leq0.4$ about the 
$\Delta\,V=0$ axis to obtain the dark gray area denoted in the lower 
panel.  The remaining objects situated between 0.0 and $-1.0$ in 
$\Delta\,V$ (denoted by the light gray shaded area) should, therefore, 
approximate the number of binaries in the cluster.  The resulting binary 
fraction can be obtained by simply summing the area indicated by the 
light gray region of the plot (447 stars) and dividing by the total area 
below the histogram between $-1.0\leq\Delta\,V\leq0.5$ (1641 stars).  
Based on this technique, our computed binary fraction for NGC~3532 is 
therefore $\sim27\%\pm5\%$, where the error represents a combination of 
Poisson statistics and uncertainties associated with the removal of the 
field star distribution.  Due to the fact, however, that a number of 
spectroscopic binaries show no appreciable brightening relative to the 
single-star main sequence \citep[see, for example, the HR diagram for 
the Hyades in Fig. 20 of][]{Perryman1998}, our derived binary frequency 
for NGC~3532 should be treated as a lower limit.

\subsection{Luminosity and Mass Functions}
\label{subsec:lfmf}

The unprecedented depth and spatial coverage of our NGC~3532 photometry 
motivates us to investigate the cluster's dynamical state via its 
luminosity and mass functions (LF and MF, respectively).  The sample of 
objects considered for this endeavor is based on the collection of cluster 
stars that have been isolated from the field population using the 
photometric filtering technique described in Section \ref{subsec:ms}.  
Constructing the LF for NGC~3532 is simply a matter of counting the number 
of main sequence stars that lie within the range of $8\leq V\leq20$ using 
a bin size of 1~mag.  To account for incompleteness in the photometry 
toward fainter magnitudes, we have added stars to the original CCD images 
in an attempt to recover them using our reduction techniques.  Briefly, 
these artificial stars were added uniformly over several trials to the 
short- and long-exposure $V$-band images for fields 1, 5, 13, 20, and 25 
(see Figure \ref{fig:fieldplot}).  The final completeness corrections to 
our luminosity function represents a combination of the results from 
these trials for all 5 fields.  The results of this exercise imply that 
our photometry remains $>99\%$ complete at the bright end (i.e., 
$V\leq15$) and $>75\%$ complete for magnitudes as faint as $V=21$. Note 
that we also estimate uncertainties in these completeness corrections 
using the techniques described in \citet{Bolte1989}.

\begin{figure}
\plotone{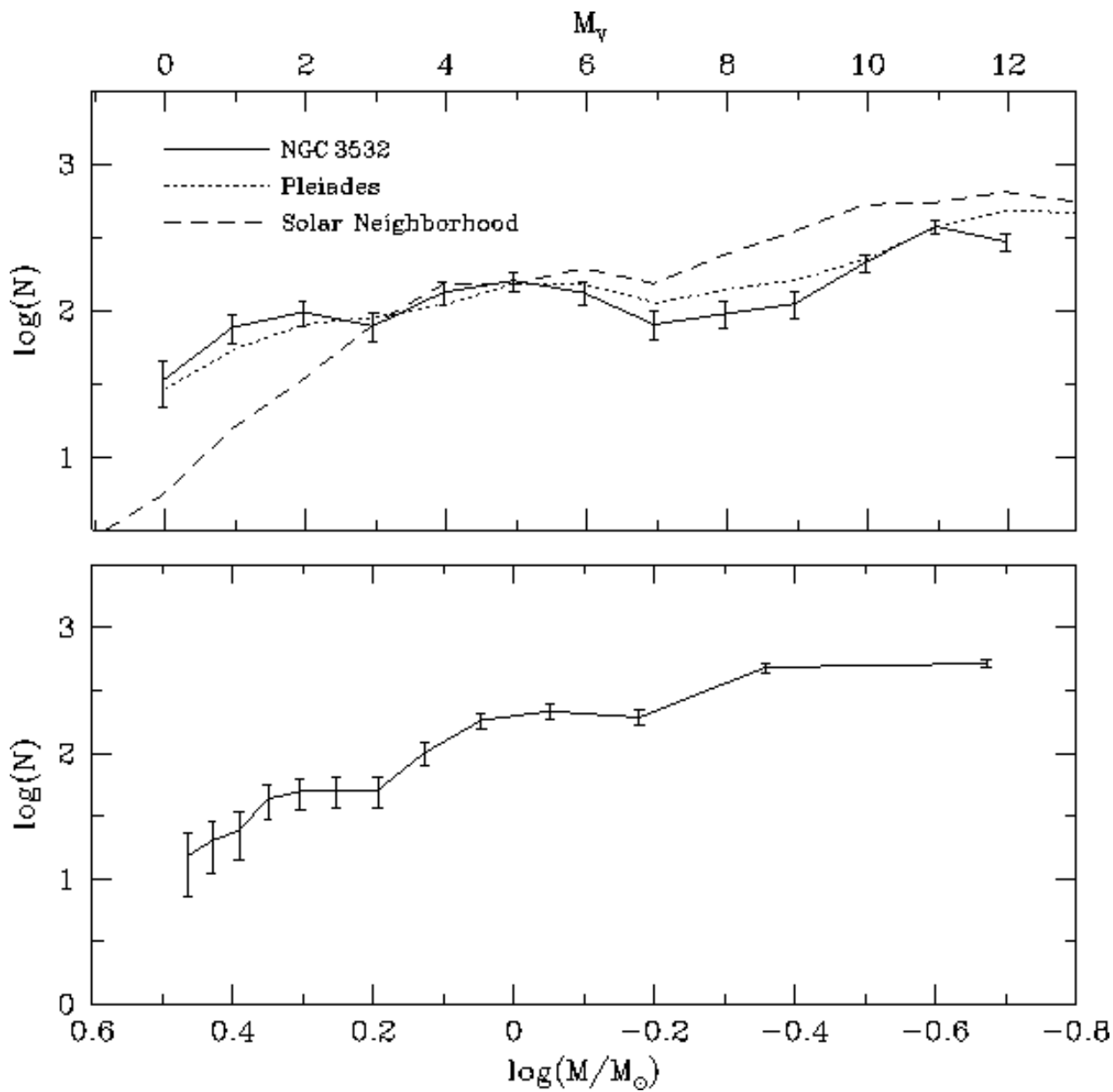}
\caption{Luminosity and mass functions for main sequence stars in 
NGC~3532 (top and bottom panels, respectively).  Error bars denote a 
combination of Poisson statistics and uncertainties in our completeness 
tests as described in the text.  The decrease in the number of stars 
around $M_{V}=7$ is reminiscent of the same ``Wielen dip" seen in 
luminosity functions for stars in the solar neighborhood (dashed line) 
and other young open clusters such as the Pleiades (dotted line).  Note 
that these latter two LFs have been normalized to the number of stars in 
NGC~3532 at $M_{V}=5$ to avoid the effect of evolution.}
\label{fig:LFMF}
\end{figure}

The final, incompleteness-corrected LF for the cluster is shown in the 
top panel of Figure \ref{fig:LFMF} as a solid line.  The error bars for 
denote a combination of uncertainties arising from simple counting 
statistics and the errors in the completeness corrections.  Note that the 
LF for NGC~3532 exhibits a small ``bump" around $4\lesssim 
M_{V}\lesssim6$ where there is a slight overabundance of stars compared 
to adjacent magnitudes.  This feature may be due to an imperfect removal 
of field stars from our photometric filtering technique.  Indeed, 
inspection of the CMDs shown in Figure \ref{fig:cmds_cleaned} reveals 
that the field disk population crosses over the cluster main sequence 
within this magnitude range.  We argue, however, that the bump is largely 
a real feature of the LF and that the decrease in the number of stars in 
the $M_{V}=7$ magnitude bin corresponds to the so-called ``Wielen dip."  
This same type of depression, which has been attributed to a change in 
the slope of the mass-luminosity relation for stars in this magnitude 
range, can also be seen in the LFs for stars in the solar neighborhood 
\citep{Wielen1974, Wielen1983, Reid2002} and other open clusters such as 
the Pleiades \citep{LeeSung1995} and Praesepe \citep{Hambly1995}.  To 
better illustrate this, as well as compare our NGC~3532 LF to other 
stellar populations, we have overplotted the LFs for the Pleiades 
\citep{LeeSung1995} and the solar neighborhood \citep{Reid2002} in the 
top panel of Figure \ref{fig:LFMF} as dotted and dashed lines, 
respectively.

Upon integrating the LF and accounting for the handful of giant and white 
dwarf stars in the cluster, we obtain a cluster population of $\sim1900$ 
stars.  However, this estimate is predominately based on single stars and 
does not include the sizable population of binary stars ($\sim27\%$) that 
we estimate in Section \ref{subsec:binaries}.  Thus, if we account for 
binary systems, the total population of stars in NGC~3532 rises to 
$\sim2400$.

The MF for NGC~3532 can be derived by using the slope of the 
mass-luminosity relation predicted from the same 300\,Myr overshooting 
isochrone employed in Section \ref{subsec:giants} to fit the cluster's 
main sequence.  The result is shown in the bottom panel of Figure 
\ref{fig:LFMF} and covers a range in mass from $\sim0.2M_{\odot}$ (the 
limit of our photometry) to $\sim3.0M_{\odot}$ (the main-sequence 
turnoff).  The best-fit slope we derive for this mass range is 
$-1.39\pm0.14$, which corresponds quite closely to that for the solar 
neighborhood \citep[$-1.35;$][]{Salpeter1955}.  We argue, however, that 
the actual MF for NGC~3532 should be better described by a broken power 
law with a much shallower slope at the low-mass end (i.e. $\lesssim 
2M_{\odot}$) and a steeper slope for more massive stars.  Assuming this 
is the case, we find the best-fit values to be $-1.04\pm0.22$ and 
$-2.54\pm0.41$ for the low- and high-mass stars, respectively.  Such a 
drastic change in slope at the two mass extremes is likely indicative 
of a mass segregation effect within the cluster.  We note, however, 
that the lowest mass bins of the MF show a very flat distribution.  
Part of this may be due to the fact that we were forced to extrapolate 
the BaSTI isochrone beyond its lowest tabulated mass ($0.5 M_{\odot}$) 
for these two bins.  Moreover, the bolometric corrections used to 
translate the isochrone luminosity to $M_V$ could be in error by as 
much as 0.5~mag for such low-mass stars.  The combination of these two 
effects lead us to doubt the reliability of the MF at the extreme 
low-mass end.

\section{Summary} 
\label{sec:conclusions}

In this investigation we have presented the results of a large, 
accurate, and homogeneous photometric $BV(RI)_{c}$ survey of the open 
cluster NGC~3532 covering approximately one square degree on the sky.  
Due to its location near the Galactic plane, the resulting CMDs 
revealed the presence of a large number of field stars that virtually 
masked the cluster's lower main sequence.  Thanks to a merger of 
infrared $JHK_{s}$ photometry from the 2MASS catalog with our data set, 
we have been able to isolate a well-populated cluster main sequence 
that extends as faint as $V\sim21$.  Moreover, our photometric 
filtering technique has allowed us to further separate out cluster 
members from the field star population to permit an accurate 
determination of the cluster's distance and reddening.  Our findings 
support previous evidence that the cluster is fairly nearby 
($d=492^{+12}_{-11}$~pc) and exhibits a remarkably low reddening 
[$E(B-V)=0.028\pm0.006$] despite its location in the disk.  Moreover, a 
robust estimate of the cluster's age ($\sim300$\,Myr) has been derived 
by fitting the latest BaSTI model isochrones to the well-populated 
upper main sequence.

The depth of our photometry has permitted the discovery of a number of new 
objects in the faint, blue region of the $[V,~(B-V)]$ CMD, the majority of 
which we believe to belong to the cluster's white dwarf sequence.  This 
belief is confirmed by the photometric recovery of eight previously known 
white dwarf stars that were discovered spectroscopically by 
\citet{ReimersKoester1989, KoesterReimers1993, Dobbie2010}.  Moreover, a 
noticeable drop in the number of white dwarfs beyond $M_{V}\sim12$ lead us 
to assume that our photometry has extended faint enough to probe the 
termination of the white dwarf cooling sequence.  While confirmation of 
this would require additional deeper photometric observations together 
with spectroscopic followup, the age of $300\pm100$\,Myr we derive by 
fitting this termination point agrees quite well with that obtained from 
analyzing the turnoff region.

We have also made a preliminary investigation of the luminosity function, 
mass function, and binarity of the stellar population contained within 
NGC~3532.  Based on our analysis of the luminosity function for NGC~3532, 
we find a decrease in the number of stars at $M_V\sim7$.  This effect, 
known as the ``Wielen dip", can also be seen in the luminosity functions 
for stars both in the field, as well as other open clusters such as 
Praesepe and the Pleiades.  The total number of stars in NGC~3532, as 
derived by integrating the luminosity function, has been determined to be 
around 1900, but we note that our estimate does not include a sizable 
number of binary stars that comprises approximately $27\%$ of the total 
cluster population.  While the cluster's overall mass function is best fit 
with a power law that has a slope close to the Salpeter value ($-1.39$), 
we argue that it is better represented by two separate slopes for the high 
and low star stars.  Based on this argument, we find that stars with 
masses less than $\sim 2M_{\odot}$ have a much shallower slope than the 
Salpeter value, which would imply that our observations do not cover the 
full extent of the cluster, or a number of low mass members have 
evaporated from the cluster due to the effects of mass segregation.

\acknowledgements{The authors extend their thanks to the staff of the 
CTIO for their superb aid during the data acquisition process.  Thanks 
also go to K.~Hainline, who helped us take an initial first look at these 
data.  This work has made use of BaSTI web tools at 
\url{http://albione.oa-teramo.inaf.it}.  This project has been funded by 
National Science Foundation grants AST~95-28177 and AST~08-03158 to 
A.~U.~L.}

\newpage{}

\begin{deluxetable}{cccrrrrr}
\tablecaption{NGC~3532 Observational Log}
\tabletypesize{\scriptsize}
\tablewidth{0pt}
\tablehead{\colhead{UT Date}                     &
           \colhead{Standards?}                  &
           \multicolumn{4}{c}{Number of Frames}  &
           \colhead{Fields Observed}             \\
           \colhead{(yymmdd)}                    &
           \colhead{}                            &
           \colhead{$B$}                         &
           \colhead{$V$}                         &
           \colhead{$R_{c}$}                     &
           \colhead{$I_{c}$}                     &
           \colhead{}                            }
\startdata
000201 & Y &   0 &  12 &   0 &   0 &        26, 27, 28 \\
000208 & N &   0 &  12 &   0 &   0 &        26, 27, 28 \\
000219 & Y &   0 &  12 &   0 &   0 &        26, 27, 28 \\
000224 & N &   0 &  12 &   0 &   0 &        26, 27, 28 \\
000225 & N &   0 & 225 &   0 &   0 &              1-25 \\
000226 & Y &  25 &  95 &  25 &  25 &        6-9, 11-25 \\
000227 & N &   0 & 105 &   0 &   0 & 1-5, 10-13, 16-25 \\
000228 & Y &  45 &  45 &  45 &  45 &	  16-19, 21-25 \\
000229 & Y &  45 &  45 &  45 &  45 &    1-3, 10-14, 20 \\
000301 & Y &  35 &  50 &  35 &  35 &     1,2, 4-10, 15 \\
000317 & N &   0 & 225 &   0 &   0 &              1-25 \\
000318 & N &   0 & 200 &   0 &   0 &              1-25 \\
000319 & N &   0 & 175 &   0 &   0 &              1-25 \\
000320 & N &   0 & 200 &   0 &   0 &              1-25 \\
000321 & N &   0 & 210 &   0 &   0 &              1-25 \\
000322 & N &   0 &  12 &   0 &   0 &        26, 27, 28 \\
000323 & N &   0 &  12 &   0 &   0 &        26, 27, 28 \\
000420 & N &   0 &  15 &   0 &   0 &        26, 27, 28 \\
000422 & N &   0 &  15 &   0 &   0 &        26, 27, 28 \\
000426 & N &   0 &  15 &   0 &   0 &        26, 27, 28 \\
000616 & N &   0 &  14 &   0 &   0 &        26, 27, 28 \\
000713 & N &   0 &  15 &   0 &   0 &        26, 27, 28 \\
\enddata
\label{tab:table1}
\end{deluxetable}

\newpage{}

\begin{deluxetable}{rcccccccccccccccccc}
\rotate
\tablecaption{$BV(RI)_{c}$ CCD Photometry and Equatorial Coordinates for Stars in NGC~3532}
\tabletypesize{\tiny}
\tablewidth{0pt}
\tablehead{\colhead{ID}                      &
           \colhead{$x$}                     &
           \colhead{$y$}                     &
           \colhead{$B$}                     &
           \colhead{$\sigma(B)$}             &
           \colhead{$N(B)$}                  &
           \colhead{$V$}                     &
           \colhead{$\sigma(V)$}             &
           \colhead{$N(V)$}                  &
           \colhead{$R_{c}$}                 &
           \colhead{$\sigma(R_{c})$}         &
           \colhead{$N(R_{c})$}              &
           \colhead{$I_{c}$}                 &
           \colhead{$\sigma(I_{c})$}         &
           \colhead{$N(I_{c})$}              &
           \colhead{$\chi$}                  &
           \colhead{$sharp$}                 &
           \colhead{RA}                      &
           \colhead{Dec}                     \\
           \colhead{(1)}                     &
           \colhead{(2)}                     &
           \colhead{(3)}                     &
           \colhead{(4)}                     &
           \colhead{(5)}                     &
           \colhead{(6)}                     &
           \colhead{(7)}                     &
           \colhead{(8)}                     &
           \colhead{(9)}                     &
           \colhead{(10)}                    &
           \colhead{(11)}                    &
           \colhead{(12)}                    &
           \colhead{(13)}                    &
           \colhead{(14)}                    &
           \colhead{(15)}                    &
           \colhead{(16)}                    &
           \colhead{(17)}                    &
           \colhead{(18)}                    &
           \colhead{(19)}                    }
\startdata
      1  &  $-$7895.5  &   7082.3  & 19.602  & 0.0550  &  3  & 18.669  & 0.0177  &  4  & 18.085  & 0.0225  &  4  & 17.404  & 0.0183  &  4  &    0.773  & $+$0.195  & 11:09:42.97  & $-$58:14:02.1 \\
      2  &  $-$7895.1  &   6996.8  & 16.313  & 0.0281  &  4  & 15.893  & 0.0056  &  4  & 15.611  & 0.0073  &  4  & 15.216  & 0.0086  &  5  &    1.303  & $+$0.058  & 11:09:43.00  & $-$58:14:23.5 \\
      3  &  $-$7893.8  &   7149.5  & 17.600  & 0.0211  &  4  & 16.253  & 0.0059  &  5  & 15.515  & 0.0054  &  5  & 14.785  & 0.0076  &  5  &    1.328  & $+$0.026  & 11:09:42.88  & $-$58:13:45.3 \\
      4  &  $-$7893.4  &   7376.1  & 20.139  & 0.0298  &  3  & 19.106  & 0.0177  &  4  & 18.436  & 0.0135  &  4  & 17.728  & 0.0144  &  5  &    0.642  & $+$0.152  & 11:09:42.76  & $-$58:12:48.7 \\
      5  &  $-$7892.6  &   7238.9  & 20.950  & 0.1233  &  3  & 20.050  & 0.0386  &  4  & 19.425  & 0.0252  &  3  & 18.677  & 0.0560  &  4  &    0.537  & $+$0.293  & 11:09:42.80  & $-$58:13:23.0 \\
      6  &  $-$7892.1  &   6776.5  & 17.645  & 0.0102  &  4  & 16.681  & 0.0054  &  5  & 16.163  & 0.0118  &  5  & 15.597  & 0.0059  &  5  &    0.925  & $+$0.004  & 11:09:43.01  & $-$58:15:18.6 \\
      7  &  $-$7891.7  &   6831.0  & 20.455  & 0.0531  &  3  & 19.329  & 0.0184  &  3  & 18.699  & 0.0122  &  4  & 18.078  & 0.0194  &  4  &    0.578  & $-$0.027  & 11:09:42.97  & $-$58:15:04.9 \\
      8  &  $-$7890.9  &   7415.1  & 20.278  & 0.0650  &  3  & 19.323  & 0.0207  &  4  & 18.776  & 0.0159  &  4  & 18.144  & 0.0235  &  4  &    0.538  & $-$0.003  & 11:09:42.66  & $-$58:12:38.9 \\
      9  &  $-$7890.7  &   7310.9  & 20.457  & 0.0689  &  3  & 19.186  & 0.0156  &  4  & 18.410  & 0.0094  &  4  & 17.626  & 0.0143  &  4  &    0.576  & $+$0.074  & 11:09:42.70  & $-$58:13:05.0 \\
     10  &  $-$7890.4  &   6529.6  & 19.036  & 0.0263  &  3  & 17.495  & 0.0041  &  4  & 16.665  & 0.0056  &  5  & 15.870  & 0.0118  &  5  &    0.721  & $+$0.014  & 11:09:43.08  & $-$58:16:20.3 \\
\enddata
\tablenotetext{~}{(1) Sequential identification number}
\tablenotetext{~}{(2) $x$-coordinate in finding chart; increases west from RA=11:05:33 at $0.25\arcsec$ pixel$^{-1}$}
\tablenotetext{~}{(3) $y$-coordinate in finding chart; increases north from Dec=$-$58:43:48 at $0.25\arcsec$ pixel$^{-1}$}
\tablenotetext{~}{(4) Photometric $B$ magnitude}
\tablenotetext{~}{(5) Standard error of the mean $B$ magnitude}
\tablenotetext{~}{(6) Number of measurements in $B$}
\tablenotetext{~}{(7) Photometric $V$ magnitude}
\tablenotetext{~}{(8) Standard error of the mean $V$ magnitude}
\tablenotetext{~}{(9) Number of measurements in $V$}
\tablenotetext{~}{(10) Photometric $R_{c}$ magnitude}
\tablenotetext{~}{(11) Standard error of the mean $R_{c}$ magnitude}
\tablenotetext{~}{(12) Number of measurements in $R_{c}$}
\tablenotetext{~}{(13) Photometric $I_{c}$ magnitude}
\tablenotetext{~}{(14) Standard error of the mean $I_{c}$ magnitude}
\tablenotetext{~}{(15) Number of measurements in $I_{c}$}
\tablenotetext{~}{(16) Mean value of $\chi$}
\tablenotetext{~}{(17) Mean value of $sharp$}
\tablenotetext{~}{(18) Right Ascension (J2000.0)}
\tablenotetext{~}{(19) Declination (J2000.0)}
\tablenotetext{~}{Note -- Table 2 will be published in electronic format.}
\label{tab:table2}
\end{deluxetable}

\newpage{}

\begin{deluxetable}{rrrrrrrrrrrrrrrrrr}
\rotate
\tablecaption{New Photometry for Previously Studied Stars}
\tabletypesize{\tiny}
\tablewidth{0pt}
\tablehead{\colhead{WEBDA}                     &
           \colhead{ID}                        &
           \colhead{RA}                        &
           \colhead{Dec}                       &
           \colhead{$B$}                       &
           \colhead{$\sigma(B)$}               &
           \colhead{$N(B)$}                    &
           \colhead{$V$}                       &
           \colhead{$\sigma(V)$}               &
           \colhead{$N(V)$}                    &
           \colhead{$R_{c}$}                   &
           \colhead{$\sigma(R_{c})$}           &
           \colhead{$N(R_{c})$}                &
           \colhead{$I_{c}$}                   &
           \colhead{$\sigma(I_{c})$}           &
           \colhead{$N(I_{c})$}                &
	   \colhead{$\chi$}                    &
           \colhead{$sharp$}                   }
\startdata
 649  & 294887  & 11:02:25.10  & -58:45:37.4  &  8.960  & 0.0062  &      5  &  7.978  & 0.0108  &      7  &  7.395  & 0.0366  &      3  &  6.906  & 0.0093  &      3  &  3.684  & -0.007 \\ 
 726  & 286900  & 11:02:41.85  & -58:58:03.3  & 18.487  & 0.0101  &      4  & 17.559  & 0.0037  &     10  & 16.975  & 0.0036  &      5  & 16.389  & 0.0086  &      5  &  0.576  &  0.008 \\ 
 725  & 283570  & 11:02:49.14  & -58:58:08.8  & 14.018  & 0.0053  &      5  & 12.129  & 0.0034  &     10  & 11.063  & 0.0087  &      5  & 10.010  & 0.0039  &      5  &  2.632  & -0.006 \\ 
 596  & 283467  & 11:02:50.60  & -58:42:07.1  &  8.854  & 0.0070  &      5  &  7.869  & 0.0079  &      6  &  7.360  & 0.0123  &      3  &  6.925  & 0.0120  &      3  &  4.344  &  0.120 \\ 
 728  & 280521  & 11:02:55.98  & -58:59:10.4  & 16.143  & 0.0034  &      5  & 15.512  & 0.0062  &     10  & 15.102  & 0.0066  &      5  & 14.691  & 0.0037  &      5  &  0.835  & -0.026 \\ 
 539  & 279539  & 11:02:58.45  & -58:57:24.8  & 11.698  & 0.0027  &      5  & 11.383  & 0.0050  &     10  & 11.173  & 0.0069  &      5  & 10.973  & 0.0036  &      5  &  2.085  & -0.013 \\ 
 727  & 277424  & 11:03:03.23  & -58:58:15.7  & 13.786  & 0.0024  &      5  & 13.471  & 0.0029  &     10  & 13.286  & 0.0053  &      5  & 13.092  & 0.0024  &      5  &  1.294  & -0.033 \\ 
 540  & 271164  & 11:03:16.18  & -58:59:55.8  & 10.559  & 0.0033  &     10  & 10.352  & 0.0052  &     20  & 10.236  & 0.0143  &     10  & 10.095  & 0.0049  &     10  &  4.626  & -0.009 \\ 
 713  & 267576  & 11:03:25.17  & -58:35:36.7  & 14.710  & 0.0054  &      5  & 13.949  & 0.0082  &     10  & 13.514  & 0.0091  &      5  & 13.104  & 0.0058  &      5  &  1.924  &  0.068 \\ 
 473  & 265155  & 11:03:29.22  & -58:47:08.6  &  9.607  & 0.0061  &      5  &  9.201  & 0.0089  &      9  &  8.938  & 0.0103  &      4  &  8.717  & 0.0053  &      5  &  5.370  &  0.138 \\ 
\enddata
\tablenotetext{~}{Note -- Table 3 will be published in electronic format.}
\label{tab:table3}
\end{deluxetable}

\newpage{}

\begin{deluxetable}{rrrrrrrrrrrrrrr}
\rotate
\tablecaption{Previous Broadband Photometry for Stars in NGC~3532}
\tabletypesize{\tiny}
\tablewidth{0pt}
\tablehead{\colhead{ }                         &
           \multicolumn{2}{c}{K59}             &
           \multicolumn{2}{c}{B77}             &
           \multicolumn{2}{c}{FS80}            &
           \multicolumn{2}{c}{J81}             &
           \multicolumn{4}{c}{WG82}            &
           \multicolumn{2}{c}{CL88}            \\
           \cline{2-3} \cline{4-5} \cline{6-7} \cline{8-9} \cline{10-13} \cline{14-15} \\
           \colhead{WEBDA}                     &
           \colhead{$V$}                       &
           \colhead{$(B-V)$}                   &
           \colhead{$V$}                       &
           \colhead{$(B-V)$}                   &
           \colhead{$V$}                       &
           \colhead{$(B-V)$}                   &
           \colhead{$V$}                       &
           \colhead{$(B-V)$}                   &
           \colhead{$V$}                       &
           \colhead{$(B-V)$}                   &
           \colhead{$(V-R_{c})$}               &
           \colhead{$(R-I)_{c}$}               &
           \colhead{$V$}                       &
           \colhead{$(B-V)$}                   }	
\startdata
 649  & \nodata & \nodata & \nodata & \nodata & \nodata & \nodata & \nodata & \nodata & \nodata & \nodata & \nodata & \nodata &   8.01  &   1.00  \\ 
 726  & \nodata & \nodata &  17.46  &   0.58  & \nodata & \nodata & \nodata & \nodata & \nodata & \nodata & \nodata & \nodata & \nodata & \nodata \\ 
 725  & \nodata & \nodata &  12.15  &   1.76  & \nodata & \nodata & \nodata & \nodata & \nodata & \nodata & \nodata & \nodata & \nodata & \nodata \\ 
 596  & \nodata & \nodata & \nodata & \nodata & \nodata & \nodata & \nodata & \nodata & \nodata & \nodata & \nodata & \nodata &   7.93  &   0.99  \\ 
 728  & \nodata & \nodata &  15.54  &   0.84  & \nodata & \nodata & \nodata & \nodata & \nodata & \nodata & \nodata & \nodata & \nodata & \nodata \\ 
 539  & \nodata & \nodata &  11.36  &   0.30  & \nodata & \nodata & \nodata & \nodata & \nodata & \nodata & \nodata & \nodata & \nodata & \nodata \\ 
 727  & \nodata & \nodata &  13.36  &   0.34  & \nodata & \nodata & \nodata & \nodata & \nodata & \nodata & \nodata & \nodata & \nodata & \nodata \\ 
 540  & \nodata & \nodata &  10.35  &   0.23  & \nodata & \nodata & \nodata & \nodata & \nodata & \nodata & \nodata & \nodata & \nodata & \nodata \\ 
 713  & \nodata & \nodata &  13.93  &   0.81  & \nodata & \nodata & \nodata & \nodata & \nodata & \nodata & \nodata & \nodata & \nodata & \nodata \\ 
 473  &   9.25  &   0.37  & \nodata & \nodata &   9.23  &   0.41  & \nodata & \nodata & \nodata & \nodata & \nodata & \nodata & \nodata & \nodata \\ 
\enddata
\tablenotetext{~}{Note -- Table 4 will be published in electronic format.}
\label{tab:table4}
\end{deluxetable}

\newpage{}

\begin{deluxetable}{llrrr}
\tablecaption{Comparisons with Previous Broadband Photometry}
\tabletypesize{\scriptsize}
\tablewidth{0pt}
\tablehead{\colhead{Data Set}               &
           \colhead{Index}                  &
           \colhead{Clipped Mean}           &
           \colhead{$N$}                    &
           \colhead{$N_{rej}$}              }
\startdata
\citet{Koelbloed1959}           & $V$         & $+0.024\pm0.042$ &  80 &  2 \\
                                & $(B-V)$     & $-0.026\pm0.031$ &  77 &  2 \\ \hline
\citet{Butler1977} ($V<16$)     & $V$         & $+0.004\pm0.018$ &  14 &  1 \\
                                & $(B-V)$     & $-0.001\pm0.080$ &  15 &  0 \\ \hline
\citet{FernandezSalgado1980}    & $V$         & $+0.029\pm0.029$ & 173 &  7 \\
                                & $(B-V)$     & $-0.002\pm0.026$ & 173 &  3 \\ \hline
\citet{Johansson1981}           & $V$         & $+0.002\pm0.048$ &  14 &  0 \\
                                & $(B-V)$     & $+0.002\pm0.020$ &  14 &  0 \\ \hline
\citet{WizinowichGarrison1982}  & $V$         & $+0.067\pm0.024$ &  68 &  0 \\ 
                                & $(B-V)$     & $-0.028\pm0.021$ &  64 &  0 \\
                                & $(V-R_{c})$ & $+0.026\pm0.030$ &  68 &  0 \\
                                & $(R-I)_{c}$ & $+0.018\pm0.023$ &  66 &  0 \\ \hline
\citet{ClariaLapasset1988}	& $V$         & $+0.058\pm0.028$ &  12 &  0 \\
                                & $(B-V)$     & $-0.006\pm0.022$ &  12 &  0 \\
\enddata
\label{tab:table5}
\end{deluxetable}

\newpage{}

\begin{deluxetable}{lccccl}
\tablecaption{Review of Parameter Estimates for NGC~3532}
\tabletypesize{\scriptsize}
\tablewidth{0pt}
\tablehead{\colhead{Reference}               &
           \colhead{$(m-M)_{0}$}             &
           \colhead{$E(B-V)$}                &
           \colhead{Age (Myr)}               &
           \colhead{[Fe/H]}                  &
           \colhead{Methods}                 }
\tablenotetext{a}{Converted from $E(b-y)$ assuming $E(B-V)\approx1.35E(b-y)$}
\startdata
\citet{Koelbloed1959}        & $8.18\pm0.20$ & $0.01$                           & 100     & \nodata          & $UBV$ photometry                 \\
\citet{FernandezSalgado1980} & $8.45\pm0.27$ & $0.042\pm0.016$                  & 200     & \nodata          & $UBV$ photometry                 \\
\citet{Eggen1981}            & $8.50\pm0.25$ & $0.031\pm0.019$\tablenotemark{a} & 350     & \nodata          & $uvbyH\beta$ photometry          \\
\citet{Johansson1981}        & $8.06\pm0.51$ & $0.10\pm0.04$                    & 200     & \nodata          & $UBV/uvbyH\beta$ photometry	\\
\citet{Schneider1987}        & \nodata       & $0.034$\tablenotemark{a}         & \nodata & \nodata          & $uvby$ photometry                \\
\citet{ClariaLapasset1988}   & \nodata       & $0.07\pm0.02$                    & \nodata & \nodata          & $UBV/DDO$ photometry             \\
\citet{Meynet1993}           & $8.35$        & $0.04$                           & 316     & \nodata          & $BV$ photometry                  \\
\citet{Luck1994}             & \nodata       & \nodata                          & \nodata & $+0.07\pm0.06$   & spectroscopy                     \\
\citet{Piatti1995}           & \nodata       & \nodata                          & \nodata & $-0.10\pm0.09$   & $DDO$ photometry                 \\
\citet{Malysheva1997}        & $8.48$        & $0.039$\tablenotemark{a}         & 229     & \nodata          & $uvbyH\beta$ photometry          \\
\citet{Twarog1997}           & $8.38$        & $0.04$                           & \nodata & $-0.02\pm0.04$   & $UBV/DDO$ photometry             \\
\citet{Robichon1999}         & $8.04\pm0.35$ & \nodata                          & \nodata & \nodata          & Hipparcos parallax               \\
\citet{Gratton2000}          & \nodata       & \nodata                          & \nodata & $+0.02\pm0.06$   & spectroscopy                     \\
\citet{Loktin2001}           & $8.43$        & $0.037$                          & 310     & \nodata          & $UBV$ photometry                 \\
\citet{Sarajedini2004}       & $8.47\pm0.07$ & \nodata                          & \nodata & \nodata          & $(V-K_{s})$ photometry           \\
\citet{Kharchenko2005}       & $8.48$        & $0.04$                           & 282     & \nodata          & photometric/kinematic            \\
\citet{vanLeeuwen2009}       & $8.07\pm0.22$ & \nodata                          & \nodata & \nodata          & Hipparcos parallax               \\
WEBDA                        & $8.43$        & $0.037$                          & 310     & $-0.02$          & various                          \\
Current Study                & $8.46\pm0.05$ & $0.028\pm0.006$                  & 300     & \nodata          & $BV(RI)_{c}~\&~JHK_{s}$ photometry   \\
\enddata
\label{tab:table6}
\end{deluxetable}

\newpage{}

\begin{deluxetable}{rrrrcccccrcrcrcc}
\rotate
\tablecaption{Photometry and Kinematic Data for Giant Stars}
\tabletypesize{\scriptsize}
\tablewidth{0pt}
\tablehead{\colhead{WEBDA ID}                     &
           \colhead{ID}                           &
           \colhead{RA}                           &
           \colhead{Dec}                          &
           \colhead{$V$}                          &
           \colhead{$(U-B)$}                      &
           \colhead{$(B-V)$}                      &
           \colhead{$(V-I_{c})$}                  &
           \colhead{$(V-K_{s})$}                  &
           \colhead{$\mu_{\alpha}$}               &
           \colhead{$\sigma(\mu_{\alpha})$}       &
           \colhead{$\mu_{\delta}$}               &
           \colhead{$\sigma(\mu_{\delta})$}       &
           \colhead{$V_{r}$}                      &
           \colhead{$\sigma(V_{r})$}              &
           \colhead{Member?}                      }
\startdata
221 &  151415 & 11:06:29.29 & $-$58:40:30.1 & 6.001 & \nodata & 1.346 & 1.167 & 2.712 &  $-$9.7 &   1.3 &  $+$5.8 &   1.3 &   $+$3.58 &    0.12 & Y \\
670 &   79390 & 11:07:57.36 & $-$58:17:26.3 & 6.978 &   1.408 & 1.372 & 1.221 & 2.817 &  $-$7.9 &   1.4 &  $+$6.5 &   1.3 &   $+$3.97 &    0.12 & Y \\
100 &  170771 & 11:06:03.84 & $-$58:41:15.9 & 7.457 &   0.930 & 1.120 & 1.033 & 2.412 &  $-$8.8 &   1.8 &  $+$5.7 &   1.8 &   $+$4.49 &    0.12 & Y \\
522 &  241462 & 11:04:13.94 & $-$58:27:51.1 & 7.590 &   0.668 & 1.232 & 1.379 & 3.225 & $-$13.8 &   1.3 &  $+$5.7 &   1.3 &  $-$21.79 &    3.13 & N \\
160 &  228568 & 11:04:35.96 & $-$58:45:20.9 & 7.624 &   0.770 & 1.027 & 0.952 & 2.257 &  $-$8.6 &   1.4 &  $+$5.8 &   1.4 &   $+$4.27 &    0.05 & Y \\
 19 &  174489 & 11:05:58.74 & $-$58:43:29.4 & 7.702 &   0.670 & 0.975 & 0.955 & 2.243 & $-$10.8 &   2.4 &  $+$3.6 &   2.2 &   $+$2.94 &    0.14 & Y \\
152 &  229744 & 11:04:33.84 & $-$58:41:39.4 & 7.751 &   0.600 & 0.925 & 0.923 & 2.123 &  $-$9.9 &   1.3 &  $+$6.7 &   1.5 &   $+$5.89 &    0.07 & Y \\
623 &   56698 & 11:08:28.81 & $-$58:53:19.6 & 7.876 &   0.220 & 0.380 & 0.550 & 1.275 &  $-$5.7 &   1.5 &  $+$3.8 &   1.5 &   $+$0.70 &    0.40 & Y \\ 
596 &  283467 & 11:02:50.60 & $-$58:42:07.1 & 7.869 &   0.686 & 0.985 & 0.944 & 2.220 &  $-$9.2 &   1.8 &  $+$5.5 &   1.9 &   $+$2.50 &    0.13 & Y \\
157 &  220113 & 11:04:50.13 & $-$58:43:03.6 & 7.901 &   1.930 & 1.647 & 1.676 & 3.564 &  $-$9.1 &   1.7 &  $-$8.3 &   1.7 &   \nodata & \nodata & N \\
273 &  241779 & 11:04:12.76 & $-$58:43:42.9 & 7.906 &   0.440 & 0.563 & 0.793 & 1.711 & $-$10.1 &   3.2 &  $+$3.8 &   3.1 &  $-$23.05 &    0.46 & N \\
649 &  294887 & 11:02:25.10 & $-$58:45:37.4 & 7.978 &   0.777 & 0.982 & 1.072 & 2.429 & $-$37.1 &   1.8 &  $-$4.5 &   1.5 &   $-$6.76 &    0.13 & N \\
122 &  183628 & 11:05:45.63 & $-$58:40:39.5 & 8.189 &   0.670 & 0.941 & 1.007 & 2.256 &  $-$9.3 &   1.8 &  $+$5.6 &   1.7 &   $+$3.34 &    0.14 & Y \\
236 &  140366 & 11:06:43.61 & $-$58:35:04.8 & 8.240 &   1.800 & 1.593 & 1.604 & 3.668 & $-$10.3 &   1.7 &  $-$4.9 &   1.7 &   \nodata & \nodata & N \\
\enddata
\label{tab:table7}
\end{deluxetable}

\clearpage{}

\clearpage

\clearpage{}

\clearpage{}

\clearpage{}

\clearpage{}

\clearpage{}

\clearpage{}

\clearpage{}

\clearpage{}

\clearpage{}

\clearpage{}

\clearpage{}

\clearpage{}

\clearpage{}

\clearpage{}

\clearpage{}

\clearpage{} 

\clearpage{}

\clearpage{}

\clearpage{}

\end{document}